\documentclass[11pt,english]{article}
\usepackage[latin9]{inputenc}
\usepackage{array}
\usepackage{mathtools}
\usepackage{url}
\usepackage{multirow}
\usepackage{amsmath}
\usepackage{amsthm}
\usepackage{amssymb}
\usepackage[letterpaper]{geometry}
\usepackage{setspace}
\usepackage[authoryear]{natbib}
\makeatletter

\providecommand{\tabularnewline}{\\}

\theoremstyle{plain}
\newtheorem{assumption}{\protect\assumptionname}
\theoremstyle{remark}
\newtheorem{rem}{\protect\remarkname}

\usepackage{amsfonts}
\usepackage{amsthm}
\usepackage{fancyhdr}
\usepackage{lastpage}
\usepackage{extramarks}
\usepackage{chngpage}
\usepackage{soul}
\usepackage{color}
\usepackage{graphics}
\usepackage{lscape}
\usepackage{url}

\usepackage{bm}
\usepackage{enumerate}
\usepackage{verbatim}
\usepackage{babel}
\usepackage{titlesec}

\def\ci{\perp\!\!\!\perp}

\usepackage{pifont}

\titleformat{\subsection}
  {\normalfont\large\bfseries}{\thesubsection}{1em}{}

\usepackage{pifont}
\usepackage{pgf}
\usepackage{tikz}\usetikzlibrary{positioning}
\usetikzlibrary{arrows}

\allowdisplaybreaks

\pdfmapfile{=Cochineal.map}

\makeatother

\usepackage{babel}
\providecommand{\assumptionname}{Assumption}
\providecommand{\remarkname}{Remark}

\begin{document}
\title{Marginal Interventional Effects\thanks{\protect\normalsize Direct all correspondence to Xiang Zhou, Department
of Sociology, Harvard University, 33 Kirkland Street, Cambridge MA
02138; email: xiang\_zhou@fas.harvard.edu}}
\author{Xiang Zhou and Aleksei Opacic\\
 Harvard University}
\maketitle
\begin{abstract}
\noindent\onehalfspacing Conventional causal estimands, such as
the average treatment effect (ATE), capture how the mean outcome in
a population or subpopulation would change if all units were assigned
to treatment versus control. Real-world policy changes, however, are
often incremental, changing treatment status for only a small segment
of the population---those at or near the \textquotedblleft margin
of participation.\textquotedblright{} To formalize this idea, two
parallel literatures in economics and in statistics and epidemiology
have developed what we call interventional effects. In this article,
we unify these perspectives by defining the interventional effect
(IE) as the per capita effect of a treatment intervention on an outcome
of interest, and the marginal interventional effect (MIE) as its limit
when the intervention size approaches zero. The IE and MIE can be
viewed as unconditional counterparts of the policy-relevant treatment
effect (PRTE) and marginal PRTE (MPRTE) from the economics literature.
Unlike the PRTE and MPRTE, however, the IE and MIE are defined without
reliance on a latent index model and can be identified either under
unconfoundedness or with instrumental variables. For both scenarios,
we show that MIEs are typically identified without the strong positivity
assumption required of the ATE, highlight several \textquotedblleft stylized
interventions\textquotedblright{} that may be particularly relevant
for policy analysis, discuss several parametric and semiparametric
estimation strategies, and illustrate the proposed methods with an
empirical example.

\noindent\clearpage{}
\end{abstract}
\sloppy\doublespacing

\section{Introduction}

Conventional causal estimands, such as the average treatment effect
(ATE) and the average treatment effect for the treated (ATT), describe
how the mean outcome in a population or subpopulation would change
if all units were assigned to treatment versus control. However, real-world
policy changes are rarely universal; they often affect only a subset
of the population---those at or near the \textquotedblleft margin
of participation\textquotedblright{} (\citealt{PRTE2001,xie2013population}).
In such cases, the ATE or ATT may misrepresent the impact of an intervention.
For instance, a college outreach program may induce only a fraction
of eligible high-school graduates to attend college. If these newly
induced students benefit substantially more than the average high-school
graduate or the average college-goer, the ATE or ATT will understate
the impact of the intervention among those actually affected, making
them a poor guide for policy evaluation.

To better mirror real-world policy changes, a growing body of research
in statistics and epidemiology has introduced the notion of interventional
effects (e.g., \citealt{murphy2001marginal,robins2004effects,taubman2009intervening,diaz2012population,diaz2013assessing,moore2012causal,haneuse2013estimation,young2014identification,kennedy2019nonparametric,naimi2020incremental,diaz2023nonparametric,wen2023intervention,diaz2024causal};
see also \citealt{korb2004varieties,shpitser2006identification,eberhardt2007interventions,tian2012identifying}
from a computer science perspective). Unlike conventional causal estimands,
interventional effects characterize how the mean outcome in the population
would respond to a hypothetical, and often incremental, change in
the treatment assignment mechanism. 

Interventions can be deterministic or stochastic. For example, \citet{taubman2009intervening}
evaluate how the 20-year risk of coronary heart disease in a cohort
of US nurses would have changed if everyone exercising fewer than
30 minutes a day had increased their exercise time to 30 minutes a
day. This intervention is deterministic in that it sets each unit's
treatment level to be $\max(30,A)$, where $A$ denotes the current
treatment level. By contrast, \citet{diaz2012population} consider
stochastic interventions where each unit's treatment level is randomly
drawn from a location-shifted treatment distribution. More recently,
for binary treatments, \citet{kennedy2019nonparametric} proposes
an incremental propensity score intervention (IPSI) that preserves
the relative odds of treatment across units with different pretreatment
characteristics. In addition to its affinity to real-world policy
shifts, a practical advantage of Kennedy's IPSI over conventional
causal estimands is that its effect can be identified without the
positivity assumption typically required to identify the ATE.

A parallel line of research in economics has addressed similar questions.
\citet{PRTE2001,HeckmanVytlacil2005} define the policy-relevant treatment
effect (PRTE) as the per capita effect of a policy change on the mean
outcome conditional on a set of pretreatment covariates. Under a latent
index model for treatment, the PRTE can be expressed as a weighted
average of the \textit{marginal treatment effect} (MTE), with weights
determined by the propensity score distributions before and after
the policy change. Building on this framework, \citet{Carneiro2010a,Carneiro2011}
define the \textit{marginal PRTE} (MPRTE) as a directional limit of
the PRTE as the alternative policy converges to the baseline. \citet{zhou2019marginal,zhou2020heterogeneous}
further modify Carneiro et al.'s approach, recasting the PRTE and
MPRTE parameters as the per capita effect of a policy change and its
limit, respectively, conditional on the propensity score rather than
the full vector of covariates. In this approach, a policy change is
allowed to vary in intensity across units with different baseline
propensities.

In this article, we unify the estimands proposed in \citet{PRTE2001,HeckmanVytlacil2005},
\citet{Carneiro2010a,Carneiro2011}, \citet{kennedy2019nonparametric},
and \citet{zhou2019marginal,zhou2020heterogeneous} in the context
of a binary treatment. We define the interventional effect (IE) and
marginal interventional effect (MIE) as the unconditional counterparts
of the PRTE and MPRTE, respectively. Unlike the PRTE and MPRTE, the
IE and MIE do not condition on covariates or the propensity score,
making them more flexible: they can accommodate interventions whose
intensity varies across individuals with different baseline characteristics.
The IE can also be interpreted as a per-capita analogue of interventional
effects studied in statistics and epidemiology. Importantly, both
the IE and MIE are defined without imposing specific causal assumptions;
they can be identified either under unconfoundedness or through instrumental
variables.

Under unconfoundedness, we show that both the IE and MIE are identified
as weighted averages of the conditional average treatment effect (CATE)
given pretreatment covariates, with weights proportional to either
the increment of the propensity score under the intervention (for
the IE) or its local derivative (for the MIE). In particular, Kennedy\textquoteright s
IPSI is a special case of the IE. Its corresponding MIE coincides
with the average treatment effect for the overlap population (ATO)
introduced by \citet{li2018balancing}, which is also akin to the
optimally weighted average treatment effect (OWATE) discussed in Crump
et al. (2006). Although the statistical properties of this estimand
are well understood, its substantive interpretation has remained somewhat
elusive; for example, \citet{li2018balancing} describe it as the
average treatment effect among \textquotedblleft the subpopulation
that currently receives either treatment in substantial proportions\textquotedblright{}
(p. 391). Our work enriches the scientific content of the ATO by showing
it to be a limit of Kennedy's IPSI---that is, the per capita effect
of an infinitesimal intervention that preserves the relative odds
of treatment across units with different pretreatment characteristics.
Moreover, we show that under unconfoundedness, conventional causal
estimands such as the ATE and ATT can also be regarded as special
cases of the MIE. 

We then extend our analysis beyond unconfoundedness to discuss the
identification of the IE and MIE using instrumental variables. Following
\citet{PRTE2001,HeckmanVytlacil2005}, we consider a latent index
model with at least one continuous instrument. In this framework,
IE can be identified as a weighted average of the MTE, provided that
the conditional support of the baseline propensity score contains
the conditional support of the propensity score under the intervention,
akin to the case of the PRTE. However, as we show, support conditions
are not required for identifying the MIE. In cases where the propensity
score under intervention depends solely on the baseline propensity
score, the IE and MIE coincide with Zhou and Xie's (2019, 2020) modified
PRTE and MPRTE; nonetheless, our identification formula for the MIE
suggests a simpler estimator than those proposed in earlier work.
Finally, in the absence of unobserved selection, the MTE reduces to
the CATE, and the identification formulas for the IE and MIE reduce
to those derived under unconfoundedness.

The rest of the paper is organized as follows. In Section 2, we introduce
the concepts of IE and MIE, outline a minimal set of assumptions for
their definition, and explicate their relationships with the PRTE
and MPRTE. In Section 3, we discuss the identification and estimation
of the MIE under unconfoundedness, highlight several special cases
and their connections to existing causal estimands, and illustrate
them by reanalyzing a dataset on the clinical effects of right heart
catheterization. In Section 4, we consider identification and estimation
of the MIE using instrumental variables, illustrating this approach
with a study on the economic returns to college. Section 5 concludes
the paper.

\section{Estimands}

\label{sec:Marginal-Interventional-Effects}

\subsection{Treatment Effects}

Let $A$ denote a binary treatment, $Y$ an outcome of interest, and
$X$ a vector of pretreatment covariates. Throughout, we assume that
all random variables are defined on a sample space consisting of units
in a superpopulation, and that their randomness arises solely from
population heterogeneity. In other words, for a given unit $i$, $X_{i},$
$A_{i},$ and $Y_{i}$ are fixed. Expectations such as $\mathbb{E}[A_{i}]$
and $\mathbb{E}[Y_{i}]$ therefore represent averages across units
in the superpopulation; for example, $\mathbb{E}[A_{i}]=\frac{1}{N}\sum_{i}A_{i}$,
where $N$ is the population size. For notational brevity, we sometimes
omit the unit subscript unless needed for clarity.

To formalize causal effects, we adopt the potential outcomes framework.
Let $Y(a)$ denote the potential outcome associated with treatment
status $a$, so that $Y(1)$ and $Y(0)$ represent the outcomes that
would be realized if a unit was assigned to treatment and control,
respectively. This framework requires the assumption of consistency.
\begin{assumption}
Consistency: For any $a\in\{0,1\}$, $Y=Y(a)$ if $A=a$.\label{assu:Consistency}
\end{assumption}
\noindent Also known as the Stable Unit Treatment Value Assumption
(SUTVA; \citealt{rubin1986comment}), the consistency assumption states
that a unit's observed outcome equals its potential outcome under
the observed treatment status. This implies that, for a given unit
$i$, $Y_{i}(a)$ is fixed. Consistency therefore requires that (i)
there are no multiple versions of the treatment, or, if such versions
exist, they yield identical potential outcomes; and (ii) each unit\textquoteright s
potential outcomes depend only on its own treatment status, thereby
ruling out spillover effects.

Under Assumption 1, the average treatment effect (ATE) is defined
as the average difference between the two potential outcomes across
all units:
\[
\textup{ATE}=\mathbb{E}[Y(1)-Y(0)].
\]
Similarly, the average treatment effect for the treated (ATT) and
for the untreated (ATU) are defined within the treated and untreated
subpopulations:
\begin{align*}
\textup{ATT} & =\mathbb{E}[Y(1)-Y(0)|A=1],\\
\textup{ATU} & =\mathbb{E}[Y(1)-Y(0)|A=0].
\end{align*}
In addition, the conditional average treatment effect (CATE) is defined
as the average treatment effect conditional on pretreatment covariates
$X$: 
\[
\textup{CATE}(x)=\mathbb{E}[Y(1)-Y(0)|X=x].
\]
Unlike the ATE, ATT, and ATU, $\textup{CATE}(x)$ is a function of
$x$, capturing how treatment effects vary across units with different
baseline characteristics. 

The above estimands admit both descriptive and prescriptive interpretations.
Descriptively, they provide numerical or functional summaries of individual
treatment effects. For example, the ATE reflects the mean individual
treatment effect in the full population, while the ATT reflects the
mean individual treatment effect among treated units. Prescriptively,
the linearity of expectation allows the ATE to be expressed as
\[
\textup{ATE}=\mathbb{E}[Y(1)]-\mathbb{E}[Y(0)],
\]
which represents how the mean outcome in the population would change
if everyone were treated versus if everyone were untreated. Likewise,
the ATT suggests how the mean outcome in the treated subpopulation
would change if all treated units were instead untreated. These prescriptive
interpretations rely on the linearity of expectation and therefore
do not extend to other numerical summaries of individual treatment
effects, such as the median.

While prescriptive interpretations of these average effects are intuitive,
they have important limitations. First, real-world policy interventions
are rarely implemented as abrupt shifts from no treatment to universal
treatment; instead, they tend to occur incrementally. Consequently,
the counterfactual scenarios underlying the ATE and related estimands
may be far removed from realistic policy changes. For example, a college
outreach program may induce only a small fraction of students---those
at the margin---to attend college. If these newly induced students
benefit substantially more from college attendance than either the
average high school graduate or the average college-goer, then the
ATE or ATT will misrepresent the program\textquoteright s actual impact.

Second, prescriptive interpretations of the ATE and ATT implicitly
require a strong form of consistency, namely that every unit\textquoteright s
potential outcomes remain fixed not only under the status quo, but
also under large-scale changes to the treatment assignment process,
such as a shift from no treatment to universal treatment. This assumption
is highly fragile in social and economic settings. For instance, due
to changing supply in skilled labor, the economic value of a college
degree for any given individual is unlikely to remain fixed if the
proportion of college graduates increases dramatically, say from 20\%
to 80\%.

Therefore, while conventional causal estimands provide useful summaries
of individual treatment effects, they can be a poor basis for designing
and evaluating policies. To address this limitation, we turn to interventional
effects, which more closely align with the kinds of incremental policy
changes typically observed in practice.

\subsection{Interventional Effects}

Let $\mathcal{I}$ denote an intervention to the treatment assignment
process, and let $A^{\mathcal{I}}$, $Y^{\mathcal{I}}$, and $X^{\mathcal{I}}$
denote the treatment, outcome, and pretreatment covariates that would
be observed under the intervention. Throughout the paper, we restrict
attention to interventions that modify only treatment assignment,
leaving all other parts of the data-generating process unchanged.
Moreover, we focus on interventions that alter the proportion of units
receiving treatment in the population. These restrictions motivate
Assumptions 2 and 3.
\begin{assumption}
System Invariance: $X^{\mathcal{I}}=X$ and $Y^{\mathcal{I}}=Y(A^{\mathcal{I}})$.\label{assu:System-invariance}
\end{assumption}
\begin{assumption}
Change: $\mathbb{E}[A^{\mathcal{I}}]-\mathbb{E}[A]\neq0$.\label{assu:Change}
\end{assumption}
Assumption \ref{assu:System-invariance} consists of two parts. First,
it rules out feedback effects of the intervention on pretreatment
covariates, requiring that interventions influence only treatment
and posttreatment variables. Second, it states that the intervention
influences outcomes only through changes in treatment status. In other
words, for each unit, the potential outcomes $Y(0)$ and $Y(1)$ are
unaffected by the intervention. This assumption will be violated if
the intervention alters potential outcomes directly (through channels
other than $A$) or indirectly via other units\textquoteright{} treatment
status (i.e., interference). For example, in evaluating the effect
of a college outreach program on earnings, Assumption \ref{assu:System-invariance}
would fail if higher college enrollment makes the labor market more
competitive among college-goers and less competitive among non-college-goers,
thereby shifting everyone\textquoteright s potential earnings---a
general equilibrium effect (\citealt{heckman1998general}). In practice,
Assumption \ref{assu:System-invariance} is more plausible for incremental
or small-scale interventions.

Assumption \ref{assu:Change} stipulates that the intervention $\mathcal{I}$
induces a change in the overall proportion of treated units. This
does not preclude the possibility that some units are induced into
treatment ($A^{\mathcal{I}}-A=1$) while others are induced out of
treatment ($A^{\mathcal{I}}-A=-1$), provided that there is a nonzero
net change in treatment prevalence. However, for interpretability
and stable estimation, it is helpful to rule out sign indeterminacy
in this net change. Therefore, in the following pages, we focus on
interventions that shift treatment propensity in a uniform direction
across the population (uniformly expansionary or uniformly contractionary),
so that the average change in treatment is bounded away from zero
and has a determinate sign. 

Given Assumption \ref{assu:Change}, we define the \textit{interventional
effect} (IE) as the change in the mean outcome per net person shifted
into treatment:
\begin{equation}
\textup{IE}=\frac{\mathbb{E}[Y^{\mathcal{I}}]-\mathbb{E}[Y]}{\mathbb{E}[A^{\mathcal{I}}]-\mathbb{E}[A]}.\label{eq:IE}
\end{equation}
Note that under Assumptions \ref{assu:Consistency} and \ref{assu:System-invariance},
the pre-intervention and post-intervention outcomes equal $Y(A)$
and $Y(A^{\mathcal{I}})$, respectively. The IE can thus be rewritten
as:
\begin{align*}
\textup{IE} & =\frac{\mathbb{E}[Y(A^{\mathcal{I}})]-\mathbb{E}[Y(A)]}{\mathbb{E}[A^{\mathcal{I}}]-\mathbb{E}[A]}.
\end{align*}
In statistics and epidemiology, interventional effects have been defined
in the unstandardized form $\mathbb{E}[Y(A^{\mathcal{I}})]-\mathbb{E}[Y(A)]$
(e.g., \citealt{robins2004effects,diaz2012population,bonvini2023incremental}).
Compared with this unstandardized version, our definition is narrower
in scope in that it applies only to interventions that change treatment
prevalence ($\mathbb{E}[A^{\mathcal{I}}]\neq\mathbb{E}[A]$). For
example, it cannot be used to gauge the effect of an intervention
that merely redistributes treatment from one subpopulation to another
(for examples of such interventions, see \citealt{opacic2025disparity},
\citealt{zhou2023higher}).

Nonetheless, the standardized IE can be particularly useful in policy
contexts where treatment is costly and $\mathbb{E}[A^{\mathcal{I}}]-\mathbb{E}[A]$
is constrained to be small due to budgetary or logistical constraints.
For example, in the case of higher education expansion, the unstandardized
interventional effect may reach its maximum under a \textquotedblleft college-for-all\textquotedblright{}
policy ($\mathbb{E}[A^{\mathcal{I}}]=1$), but such a drastic increase
in enrollment is unlikely to be economically or politically feasible.
In such cases, policymakers may prefer to evaluate interventions based
on the per-person benefit, as captured by the standardized IE in Equation
\eqref{eq:IE}.

The IE defined in Equation \eqref{eq:IE} is closely related to \citeauthor{PRTE2001}'s
(2001) (per capita) PRTE, which is defined as the change in the mean
outcome per net person shifted into treatment, conditional on a vector
of pretreatment covariates $X$:
\[
\textup{PRTE}(x)=\frac{\mathbb{E}[Y^{\mathcal{I}}|X=x]-\mathbb{E}[Y|X=x]}{\pi^{\mathcal{I}}(x)-\pi(x)},
\]
where $\pi(x)\stackrel{\Delta}{=}\mathbb{E}[A|X=x]$ and $\pi^{\mathcal{I}}(x)\stackrel{\Delta}{=}\mathbb{E}[A^{\mathcal{I}}|X=x]$
are the pre- and post-intervention propensity scores at covariate
value $X=x$.

The distinction between IE and $\textup{PRTE}(x)$ is important. IE
is well-defined whenever the intervention shifts the overall proportion
of treated units, whereas $\textup{PRTE}(x)$ is well-defined only
at covariate values where the conditional probability of treatment
changes. Moreover, even if the intervention alters both the overall
treatment prevalence and the conditional probability at every covariate
value, $\textup{IE}\neq\mathbb{E}[\textup{PRTE}(X)]$. Instead, it
is a weighted average of $\textup{PRTE}(X)$:

\begin{equation}
\textup{IE}=\mathbb{E}\big[\underbrace{\frac{\pi^{\mathcal{I}}(X)-\pi(X)}{\mathbb{E}[\pi^{\mathcal{I}}(X)-\pi(X)]}\cdot}_{\stackrel{\Delta}{=}w_{\textup{PRTE}}^{\textup{IE}}(X)}\textup{PRTE}(X)\big].\label{eq:IE-PRTE}
\end{equation}

Equation \eqref{eq:IE-PRTE} shows that the impact of an intervention,
as summarized by the IE, depends not only on interventional effects
among units with covariate strata ($\textup{PRTE}(x)$) but also on
the relative intensity of the intervention across strata ($w_{\textup{PRTE}}^{\textup{IE}}(x)$).
For example, consider a tuition subsidy for higher education (e.g.,
delivered through tax credits). If $X$ denotes the student's family
income, then the interventional effect of the subsidy is shaped by
both the interventional effects within income groups ($\textup{PRTE}(x)$)
and the relative responsiveness to the tuition subsidy across students
from different income backgrounds ($w_{\textup{PRTE}}^{\textup{IE}}(x)$).

\subsection{Marginal Interventional Effects}

\label{subsec:Marginal-Interventional-Effects}

\noindent We now shift our focus from generic interventions to \textit{marginal}
interventions. Following \citet{Carneiro2010a}, we use the term ``marginal''
to mean an infinitesimal perturbation from the status quo. Formally,
consider a class of interventions indexed by a positive scalar $\delta$:
$\{\mathcal{I}_{\delta}:\delta>0\}$. Let $O=(X,A,Y)$ denote the
observed data under the status quo and $O^{\mathcal{I_{\delta}}}=(X^{\mathcal{I_{\delta}}},A^{\mathcal{I_{\delta}}},Y^{\mathcal{I_{\delta}}})$
the data that would be observed under intervention $\mathcal{I}_{\delta}$.
For completeness, we extend this notation by defining $O^{\mathcal{I}_{0}}=O$.
Therefore, $\mathcal{I}_{0}$ can be viewed as a null intervention
that preserves the status quo, and the effect of $\mathcal{I}_{\delta}$
can be assessed by comparing outcomes under $\mathcal{I}_{\delta}$
and $\mathcal{I}_{0}$. The IE defined previously can thus be rewritten
as
\[
\textup{IE}_{\delta}=\frac{\mathbb{E}[Y^{\mathcal{I}_{\delta}}]-\mathbb{E}[Y^{\mathcal{I}_{0}}]}{\mathbb{E}[A^{\mathcal{I}_{\delta}}]-\mathbb{E}[A^{\mathcal{I}_{0}}]}.
\]

To fix ideas, consider the following classes of stochastic interventions,
in which each unit\textquoteright s treatment status is independently
drawn from a Bernoulli distribution with probability $\pi^{\mathcal{I}_{\delta}}(x)$:
\begin{enumerate}
\item[(a)] $\pi^{\mathcal{I}_{\delta}}(x)=\min\{1,\pi(x)+\delta\}$;
\item[(b)] $\pi^{\mathcal{I}_{\delta}}(x)=\min\{1,\pi(x)e^{\delta}\}$;
\item[(c)] $\pi^{\mathcal{I}_{\delta}}(x)=1-(1-\pi(x))e^{-\delta}$. 
\end{enumerate}
Intervention (a) imposes an additive shift to the baseline propensity
score $\pi(x)$, whereas intervention (b) applies a multiplicative
shift. In intervention (c), we have $1-\pi^{\mathcal{I}_{\delta}}(x)=(1-\pi(x))e^{-\delta}$
. Thus, it represents a proportional reduction of the probability
of not receiving treatment. Intervention (a) is \textquotedblleft neutral\textquotedblright{}
in that it increases everyone's propensity score by the same additive
amount (until it reaches one), while interventions (b) and (c) generate
unequal increments that depend on each unit\textquoteright s baseline
propensity score. Consequently, when treatment effects vary with the
baseline propensity, these interventions should induce unequal effects
at the margin.

To formalize the concept of marginal interventional effects, we impose
the following continuity and differentiability assumptions.
\begin{assumption}
Continuity: \textup{$\lim_{\delta\downarrow0}\mathbb{E}[A^{\mathcal{I}_{\delta}}]=\mathbb{E}[A^{\mathcal{I}_{0}}]$
and $\lim_{\delta\downarrow0}\mathbb{E}[Y^{\mathcal{I}_{\delta}}]=\mathbb{E}[Y^{\mathcal{I}_{0}}]$}.\label{assu:Continuity}
\end{assumption}
\begin{assumption}
Differentiability: \textup{Both $\mathbb{E}[A^{\mathcal{I}_{\delta}}]$
and $\mathbb{E}[Y^{\mathcal{I}_{\delta}}]$ are continuously differentiable
on $[0,\epsilon)$ for some $\epsilon>0$, and $\frac{\partial\mathbb{E}[A^{\mathcal{I}_{\delta}}]}{\partial\delta}\big|_{\delta=0}>0$}.\label{assu:Differentiability}
\end{assumption}
Assumption \ref{assu:Continuity} requires that both the prevalence
of treatment and the mean outcome be right-continuous in $\delta$.
In the tuition subsidy example, if $\delta$ denotes the amount of
subsidy, then continuity means that as the subsidy approaches zero,
the proportion of students attending college converges to its baseline
level, and so does the average outcome (e.g., labor market earnings)
in the population. This condition is satisfied, for example, in a
latent index model of the form (\citealt{HeckmanVytlacil2005}):
\begin{align*}
A^{\mathcal{I}_{\delta}}=\mathbb{I}(\delta+g(X)-V\geq0),\quad\forall\delta\in[0,\epsilon),
\end{align*}
where $g(X)$ is a function of $X$, and $V$ is an unobserved idiosyncratic
cost of treatment. In this model, we have sure continuity, i.e., $\lim_{\delta\downarrow0}A^{\mathcal{I}_{\delta}}=A$
for all units, which in turn implies $\lim_{\delta\downarrow0}Y^{\mathcal{I}_{\delta}}=\lim_{\delta\downarrow0}Y(A^{\mathcal{I}_{\delta}})=Y(A)=Y$
for all units. These sure continuity conditions guarantee continuity
in expectation. 

Sure continuity, however, may not be necessary for Assumption \ref{assu:Continuity}
to hold. For example, in any of the stochastic interventions mentioned
above, $\lim_{\delta\downarrow0}A^{\mathcal{I}_{\delta}}\neq A$,
because treatment remains random for all units with $\pi^{\mathcal{I_{\delta}}}(x)\in(0,1)$,
no matter how small $\delta$ is. Nonetheless, $\lim_{\delta\downarrow0}\mathbb{E}[A^{\mathcal{I}_{\delta}}]=\mathbb{E}[A^{\mathcal{I}_{0}}]$
still holds because $\lim_{\delta\downarrow0}\pi^{\mathcal{I_{\delta}}}(x)=\pi(x)$
for all $x\in\textup{supp}(X)$, which implies $\lim_{\delta\downarrow0}\mathbb{E}[A^{\mathcal{I}_{\delta}}]=\mathbb{E}[A]$.
In this case, however, it is not guaranteed that $\lim_{\delta\downarrow0}\mathbb{E}[Y^{\mathcal{I}_{\delta}}]=\mathbb{E}[Y^{\mathcal{I}_{0}}]$.

In addition to Assumption \ref{assu:Continuity}, we also invoke Assumption
5 (\textit{Differentiability}) so that we can define and evaluate
the limit of $\textup{IE}_{\delta}$ as $\delta$ approaches zero
via L'Hôpital's rule. Specifically, for a class of interventions satisfying
Assumptions \ref{assu:Consistency}-\ref{assu:Differentiability},
we define the \textit{marginal interventional effect} (MIE) as
\begin{equation}
\textup{MIE}=\lim_{\delta\downarrow0}\textup{IE}_{\delta}=\lim_{\delta\downarrow0}\frac{\mathbb{E}[Y^{\mathcal{I}_{\delta}}]-\mathbb{E}[Y^{\mathcal{I}_{0}}]}{\mathbb{E}[A^{\mathcal{I_{\delta}}}]-\mathbb{E}[A^{\mathcal{I}_{0}}]}=\frac{\partial\mathbb{E}[Y^{\mathcal{I}_{\delta}}]/\partial\delta\big|_{\delta=0}}{\partial\mathbb{E}[A^{\mathcal{I}_{\delta}}]/\partial\delta\big|_{\delta=0}}.\label{eq:MIE}
\end{equation}
The MIE thus captures the per capita effect of an infinitesimal intervention
within the class $\{\mathcal{I}_{\delta}:\delta>0\}$. The MIE is
useful for evaluating the relative impact of alternative interventions
at the margin. For example, the MIE under a uniform tuition subsidy
for all students may differ from that under a means-tested financial
aid program.

The MIE is closely related to \citeauthor{Carneiro2010a}'s MPRTE,
which is defined as the limit of $\textup{PRTE}(x)$: 
\[
\textup{MPRTE}(x)=\lim_{\delta\downarrow0}\textup{PRTE}_{\delta}(x)=\lim_{\delta\downarrow0}\frac{\mathbb{E}[Y^{\mathcal{I}_{\delta}}|X=x]-\mathbb{E}[Y^{\mathcal{I}_{0}}|X=x]}{\pi^{\mathcal{I_{\delta}}}(x)-\pi^{\mathcal{I}_{0}}(x)},
\]
where $\pi^{\mathcal{I}_{0}}(x)\stackrel{\Delta}{=}\mathbb{E}[A^{\mathcal{I}_{0}}|X=x]$
and $\pi^{\mathcal{I_{\delta}}}(x)\stackrel{\Delta}{=}\mathbb{E}[A^{\mathcal{I_{\delta}}}|X=x]$
are the pre- and post-intervention propensity scores at covariate
value $x$.

Under suitable regularity conditions that allow us to exchange limits
and integration and apply L'H\^{o}pital's rule, Equations \eqref{eq:IE-PRTE}
and \eqref{eq:MIE} imply that
\begin{align}
\textup{MIE} & =\lim_{\delta\downarrow0}\frac{\mathbb{E}[Y^{\mathcal{I}_{\delta}}]-\mathbb{E}[Y^{\mathcal{I}_{0}}]}{\mathbb{E}[A^{\mathcal{I_{\delta}}}]-\mathbb{E}[A^{\mathcal{I}_{0}}]}\nonumber \\
 & =\lim_{\delta\downarrow0}\mathbb{E}\big[\frac{\pi^{\mathcal{I}_{\delta}}(X)-\pi^{\mathcal{I}_{0}}(X)}{\mathbb{E}[\pi^{\mathcal{I}_{\delta}}(X)-\pi^{\mathcal{I}_{0}}(X)]}\cdot\frac{\mathbb{E}[Y^{\mathcal{I}_{\delta}}|X]-\mathbb{E}[Y^{\mathcal{I}_{0}}|X]}{\pi^{\mathcal{I_{\delta}}}(X)-\pi^{\mathcal{I}_{0}}(X)}\big]\nonumber \\
 & =\mathbb{E}\big[\lim_{\delta\downarrow0}\frac{\pi^{\mathcal{I}_{\delta}}(X)-\pi^{\mathcal{I}_{0}}(X)}{\mathbb{E}[\pi^{\mathcal{I}_{\delta}}(X)-\pi^{\mathcal{I}_{0}}(X)]}\cdot\lim_{\delta\downarrow0}\frac{\mathbb{E}[Y^{\mathcal{I}_{\delta}}|X]-\mathbb{E}[Y^{\mathcal{I}_{0}}|X]}{\pi^{\mathcal{I_{\delta}}}(X)-\pi^{\mathcal{I}_{0}}(X)}\big]\nonumber \\
 & =\mathbb{E}\big[\lim_{\delta\downarrow0}\frac{(\pi^{\mathcal{I_{\delta}}}(X)-\pi^{\mathcal{I}_{0}}(X))}{\mathbb{E}[(\pi^{\mathcal{I_{\delta}}}(X)-\pi^{\mathcal{I}_{0}}(X))]}\cdot\textup{MPRTE}(X)\big]\nonumber \\
 & =\mathbb{E}\big[\underbrace{\frac{\dot{\pi}^{\mathcal{I}_{0}}(X)}{\mathbb{E}[\dot{\pi}^{\mathcal{I}_{0}}(X)]}}_{\stackrel{\Delta}{=}w_{\textup{MPRTE}}^{\textup{MIE}}(X)}\textup{MPRTE}(X)\big],\label{eq:MIE-MPRTE}
\end{align}
where $\dot{\pi}^{\mathcal{I}_{0}}(x)\stackrel{\Delta}{=}\partial\pi^{\mathcal{I_{\delta}}}(x)/\partial\delta|_{\delta=0}$.
Thus, the MIE is a weighted mean of $\textup{MPRTE}(X)$, with weights
proportional to the derivative of the propensity score at the status
quo. These weights capture the relative intensity of an infinitesimal
intervention across covariate strata. For example, under a means-tested
financial aid program, $w_{\textup{MPRTE}}^{\textup{MIE}}(x)$ will
tend to be greater for low-income students than under a uniform tuition
subsidy.

In this way, the MIE summarizes who at the margin is most responsive
to an intervention and how much they benefit, thereby providing a
concise measure of the policy\textquoteright s local effectiveness.
Unlike the average treatment effect (ATE) or the local average treatment
effect (LATE; \citealt{ImbensAngrist1994,AngristImbensRubin1996}),
which describe average causal responses to treatment assignment, the
MIE explicitly characterizes the impact of \textit{marginal policy
shifts}. Compared with $\textup{MPRTE}(x)$, which describes per-unit
effects conditional on covariates, the MIE aggregates these effects
across the population while weighting them by the relative intensity
of the intervention, making it particularly useful for assessing the
overall impact of small-scale program adjustments.

In an influential study of the ``marginal returns'' to college,
\citet{Carneiro2011} estimated $\textup{MPRTE}(x)$ under several
stylized policy interventions, including additive and proportional
changes in each individual\textquoteright s propensity to attend college,
conditional on background characteristics $X$ and a set of instrumental
variables such as distance to college (see Section \ref{sec:IV}).
They then evaluated the ``overall'' MPRTEs by averaging the corresponding
$\textup{MPRTE}(x)$'s over the marginal distribution of $X$, i.e.,
$\mathbb{E}[\textup{MPRTE}(X)]$. This approach effectively assumes
uniform weighting ($w_{\textup{MPRTE}}^{\textup{MIE}}(x)=1$), thereby
ruling out the possibility that an intervention may affect individuals
differently depending on their background characteristics. In reality,
interventions are often preferential or targeted: for example, a college
outreach program is likely to increase college attendance more among
low-income youth than among high-income youth. The impacts of such
interventions are therefore better captured or approximated by the
MIE, which accounts for heterogeneity in the intensity of treatment
shifts across individuals, rather than by the unweighted mean of $\textup{MPRTE}(X)$.

\section{Identification and Estimation under Unconfoundedness}

\label{sec:Unconfoundedness}

\noindent In this section, we discuss the identification and estimation
of the IE and MIE under the assumption of unconfoundedness, i.e.,
the absence of unobserved confounding in the treatment--outcome relationship
once we condition on a set of pretreatment covariates $X$. This assumption
is satisfied by design in a complete or stratified randomized experiment
(\citealt{imbens2015causal}), and in observational settings it can
sometimes be justified based on substantive domain knowledge. 

We focus on interventions that satisfy Assumptions \ref{assu:Consistency}-\ref{assu:Differentiability}.
To sharpen identification, we also assume that the derivative $\dot{\pi}^{\mathcal{I}_{0}}(x)\stackrel{\Delta}{=}\partial\pi^{\mathcal{I_{\delta}}}(x)/\partial\delta|_{\delta=0}$
exists for all $x\in\textup{supp}(X)$ and that it is known---either
directly as a function of $x$ or as a function of the baseline propensity
score $\pi(x)$ (which itself may need to be estimated). These conditions
are satisfied, for example, if $\pi^{\mathcal{I_{\delta}}}(x)$ is
a smooth function of $\delta$ and $\pi(x)$. 

\subsection{General Identification Results}

Formally, to identify the IE and MIE, we invoke the following two
assumptions.
\begin{assumption}
Unconfoundedness: For any $a\in\{0,1\}$ and any $\delta\in[0,\infty)$,
$Y(a)\ci A^{\mathcal{I}_{\delta}}|X$.\label{assu:Unconfoundedness}
\end{assumption}
\begin{assumption}
Support: For any $\delta\in[0,\infty)$, $\{x:\pi^{\mathcal{I_{\delta}}}(x)-\pi(x)\neq0\}\subset\{x:0<\pi(x)<1\}$\label{assu:Support}
\end{assumption}
Assumption \ref{assu:Unconfoundedness} states that, conditional on
covariates $X$, treatment assignment is independent of the potential
outcomes---that is, treatment is \textquotedblleft as if random\textquotedblright{}
within covariate strata. Importantly, this assumption applies not
only to counterfactual post-intervention scenarios ($\delta>0$) but
also to the pre-intervention world ($\delta=0$). Thus, even when
the intervention $\mathcal{I}_{\delta}$ is explicitly stochastic
with known assignment probabilities, violations of unconfoundedness
may still arise if the baseline data are already subject to unobserved
confounding.

Assumption \ref{assu:Support} requires the increment of the propensity
score, $\pi^{\mathcal{I_{\delta}}}(x)-\pi(x)$, be zero in any region
where treatment is deterministic under the status quo. In other words,
the intervention must not alter treatment probabilities for covariate
values where $\pi(x)$ is zero or one. This condition is weaker than
the standard positivity assumption used in identifying the ATE, which
requires that $0<\pi(x)<1$ for all $x\in\textup{supp}(X)$. Here,
we require positivity only where the intervention actually modifies
treatment probabilities.

Under Assumptions \ref{assu:Consistency}-\ref{assu:Unconfoundedness},
for every $x\in\{x:\pi^{\mathcal{I_{\delta}}}(x)-\pi(x)\neq0\}$,
we have 
\begin{align}
\textup{PRTE}_{\delta}(x) & =\frac{\mathbb{E}[Y^{\mathcal{I}_{\delta}}-Y^{\mathcal{I}_{0}}|X=x]}{\mathbb{E}[A^{\mathcal{I}_{\delta}}-A^{\mathcal{I}_{0}}|X=x]}\nonumber \\
 & =\frac{\mathbb{E}[Y(0)+A^{\mathcal{I}_{\delta}}(Y(1)-Y(0))-Y(0)-A^{\mathcal{I}_{0}}(Y(1)-Y(0))|X=x]}{\mathbb{E}[A^{\mathcal{I}_{\delta}}-A^{\mathcal{I}_{0}}|X=x]}\quad(\textup{consistency, system invariance})\nonumber \\
 & =\frac{\mathbb{E}[(A^{\mathcal{I}_{\delta}}-A^{\mathcal{I}_{0}})(Y(1)-Y(0))|X=x]}{\mathbb{E}[A^{\mathcal{I}_{\delta}}-A^{\mathcal{I}_{0}}|X=x]}\nonumber \\
 & =\frac{\mathbb{E}[A^{\mathcal{I}_{\delta}}-A^{\mathcal{I}_{0}}|X=x]\mathbb{E}[Y(1)-Y(0)|X=x]}{\mathbb{E}[A^{\mathcal{I}_{\delta}}-A^{\mathcal{I}_{0}}|X=x]}\quad(\textup{unconfoundedness})\nonumber \\
 & =\textup{CATE}(x).\label{eq:PRTE-CATE}
\end{align}
Thus, by Equation \eqref{eq:IE-PRTE}, the IE associated with intervention
$\mathcal{I}_{\delta}$ is a weighted mean of $\textup{CATE}(X)$:
\begin{equation}
\textup{IE}_{\delta}=\mathbb{E}\big[\underbrace{\frac{\pi^{\mathcal{I_{\delta}}}(X)-\pi^{\mathcal{I}_{0}}(X)}{\mathbb{E}[\pi^{\mathcal{I_{\delta}}}(X)-\pi^{\mathcal{I}_{0}}(X)]}}_{\stackrel{\Delta}{=}w_{\textup{CATE}}^{\textup{IE}_{\delta}}(X)}\cdot\textup{CATE}(X)\big].\label{eq:IE-CATE}
\end{equation}
The CATE, in turn, is identified as the conditional mean difference
given $X$, that is, $\mathbb{E}[Y|X,A=1]-\mathbb{E}[Y|X,A=0]$. From
Equation \eqref{eq:IE-CATE}, it follows that $\textup{IE}_{\delta}$
is identified as long as $\textup{CATE}(x)$ is identified on $\{x:\pi^{\mathcal{I_{\delta}}}(x)-\pi^{\mathcal{I}_{0}}(x)\neq0\}$.
Unlike the ATE, it does not require $\textup{CATE}(x)$ to be identified
over the full support of $X$.

Given Equation \eqref{eq:MIE-MPRTE} and the fact that $\textup{MPRTE}(x)=\lim_{\delta\downarrow0}\textup{PRTE}_{\delta}(x)=\textup{CATE}(x)$,
the MIE can likewise be written as a weighted mean of $\textup{CATE}(X)$
\begin{equation}
\textup{MIE}=\mathbb{E}\big[\underbrace{\frac{\dot{\pi}^{\mathcal{I}_{0}}(X)}{\mathbb{E}[\dot{\pi}^{\mathcal{I}_{0}}(X)]}}_{\stackrel{\Delta}{=}w_{\textup{CATE}}^{\textup{MIE}}(X)}\textup{CATE}(X)\big].\label{eq:MIE-CATE}
\end{equation}
Thus, for a given class of interventions, the MIE weight $w_{\textup{CATE}}^{\textup{MIE}}(x)$
is proportional to $\dot{\pi}^{\mathcal{I}_{0}}(X)$, which gauges
the relative intensity of an infinitesimal intervention at covariate
values $x$.
\begin{rem}
From Equations \eqref{eq:IE-CATE} and \eqref{eq:MIE-CATE}, we can
see that the denominators of the weights for the IE and MIE correspond
to the net change in treatment prevalence---at the finite level for
the IE and at the margin for the MIE. For these quantities to be well-defined,
their sign must be determinate. This condition will be automatically
satisfied when the intervention of interest shifts treatment propensity
in the same direction across the population, i.e., if it is either
uniformly expansionary or uniformly contractionary. By contrast, interventions
with a redistributive component---expanding treatment for some subgroups
while contracting it for others---may induce sign uncertainty. For
example, if $\pi^{\mathcal{I_{\delta}}}(x)$ is a function of the
baseline propensity score $\pi(x)$, then estimation uncertainty in
$\hat{\pi}(x)$ may lead to ambiguity in the sign of the estimated
IE and MIE. Therefore, throughout the rest of the paper, we focus
on interventions that are uniformly directional to ensure interpretability
and stable estimation.
\end{rem}

\subsection{Stylized Interventions}

\label{subsec:Special-Cases}

\noindent From Equations \eqref{eq:IE-CATE} and \eqref{eq:MIE-CATE},
we can see that $\textup{IE}_{\delta}$ and MIE depend on $\mathcal{I}_{\delta}$
only through the modified propensity score $\pi^{\mathcal{I}_{\delta}}(X)$
and its local derivative with respect to $\delta$. In other words,
two different classes of interventions inducing the same form of $\pi^{\mathcal{I}_{\delta}}(X)$
will yield the same values of $\textup{IE}_{\delta}$ and MIE. Thus,
we may focus on interventions defined up to their induced propensity
score $\pi^{\mathcal{I}_{\delta}}(X)$. Below, we consider several
special cases where $\pi^{\mathcal{I}_{\delta}}(X)$ depends on $X$
only through the baseline propensity score $\pi(X)$.

We begin with the three stylized interventions introduced in Section
\ref{subsec:Marginal-Interventional-Effects} (see also \citealt{zhou2020heterogeneous}):
\begin{enumerate}
\item[(a)] $\pi^{\mathcal{I}_{\delta}}(x)=\min\{1,\pi(x)+\delta\}$;
\item[(b)] $\pi^{\mathcal{I}_{\delta}}(x)=\min\{1,\pi(x)e^{\delta}\}$;
\item[(c)] $\pi^{\mathcal{I}_{\delta}}(x)=1-(1-\pi(x))e^{-\delta}$. 
\end{enumerate}
Intervention (a) is ``neutral'' in that it raises everyone's propensity
score by the same additive amount (until it reaches one). In this
case, $w_{\textup{CATE}}^{\textup{MIE}}(x)\propto\dot{\pi}^{\mathcal{I}_{0}}(x)=1$
whenever $\pi(x)<1$, and $w_{\textup{CATE}}^{\textup{MIE}}(x)=0$
otherwise. Hence, the corresponding MIE is an unweighted mean of $\textup{CATE}(X)$
over the region $\{x:0\leq\pi(x)<1\}$, which coincides with ATE if
$\textup{supp}(\pi(X))\subset[0,1)$. Intervention (b) scales everyone's
propensity score multiplicatively (again capped at one). Because the
increment grows with $\pi(x)$, this intervention is ``disequalizing'':
those who are already more likely to be treated under the status quo
receive a larger boost. Under this intervention, $w_{\textup{CATE}}^{\textup{MIE}}(x)\propto\dot{\pi}^{\mathcal{I}_{0}}(x)=\pi(x)$
when $\pi(x)<1$; otherwise $w_{\textup{CATE}}^{\textup{MIE}}(x)=0$.
The resulting MIE is a weighted mean of $\textup{CATE}(X)$ with weights
proportional to the baseline propensity score. If $\textup{supp}(\pi(X))\subset[0,1)$,
this estimand is equal to the ATT. By contrast, intervention (c) is
``equalizing.'' Its derivative satisfies $\dot{\pi}^{\mathcal{I}_{0}}(x)=1-\pi(x)$,
so units with a lower baseline propensity score receive a larger increment.
The corresponding MIE is again a weighted mean of $\textup{CATE}(X)$,
this time with weights proportional to $1-\pi(x)$. This estimand,
not surprisingly, is equal to the ATU. 

Closely related to intervention (c), \citet{wen2023intervention}
proposed a ``multiplicative shift'' intervention, defined as $1-\pi^{\mathcal{I}_{\delta}}(x)=(1-\delta)(1-\pi(x)),$
where $1-\delta$ represents a factor by which the probability of
non-treatment is reduced (Here we redefine Wen et al.'s original $\delta$
as $1-\delta$ so that $\delta=0$ corresponds to the status quo).
At the margin, this intervention is equivalent to (c), since in both
cases $\dot{\pi}^{\mathcal{I}_{0}}(x)=1-\pi(x)$. Similarly, \citet{diaz2024causal}
considered the mirror image of Wen et al.\textquoteright s intervention,
defined as $\pi^{\mathcal{I}_{\delta}}(x)=(1+\delta)\pi(x)$, where
$1+\delta$ represents a factor by which the probability of treatment
is increased. At the margin, \citeauthor{diaz2024causal}'s intervention
is equivalent to (b), since in both cases $\dot{\pi}^{\mathcal{I}_{0}}(x)=\pi(x)$.

The stylized interventions (a)-(c) are summarized in the first three
rows of Table \ref{tab:Four-stylized-interventions}. Note that the
MIE is identified if and only if $\{x:\dot{\pi}^{\mathcal{I}_{0}}(x)\neq0\}\subset\{x:0<\pi(x)<1\}$.
This condition implies that intervention (b) is always identified,
whereas interventions (a) and (c) are identified only when the baseline
propensity score is strictly positive for all units ($\textup{supp}(\pi(X))\subset(0,1]$).

\begin{table}
\caption{Four stylized interventions and the associated marginal interventional
effects.\label{tab:Four-stylized-interventions}}

\begin{centering}
{\small{}%
\begin{tabular}{lc>{\centering}m{3.75cm}>{\centering}m{3.75cm}}
\hline 
{\small$\pi^{\mathcal{I}_{\delta}}(X)$} & {\small MIE weight ($\propto\dot{\pi}^{\mathcal{I}_{0}}(X)$)} & {\small Connection to existing estimands} & {\small Positivity required for identification?}\tabularnewline
\hline 
{\small$\min\{1,\pi(x)+\delta\}$} & {\small$\propto\mathbb{I}(\pi(x)<1)$} & {\small ATE if $\textup{supp}(\pi(X))\subset[0,1)$} & {\small Partially ($\textup{supp}(\pi(X))\subset(0,1]$)}\tabularnewline
{\small$\min\{1,\pi(x)e^{\delta}\}$} & {\small$\propto\pi(x)\mathbb{I}(\pi(x)<1)$} & {\small ATT if $\textup{supp}(\pi(X))\subset[0,1)$} & {\small No}\tabularnewline
{\small$1-(1-\pi(x))e^{-\delta}$} & {\small$\propto1-\pi(x)$} & {\small ATU} & {\small Partially ($\textup{supp}(\pi(X))\subset(0,1]$)}\tabularnewline
{\small$\frac{e^{\delta}\pi(x)}{1-\pi(x)+e^{\delta}\pi(x)}$} & {\small$\propto\pi(x)(1-\pi(x))$} & {\small ATO} & {\small No}\tabularnewline
\hline 
\end{tabular}}{\small\smallskip{}
}{\small\par}
\par\end{centering}
Note: ATE = average treatment effect; ATT = average treatment effect
for the treated; ATU = average treatment effect for the untreated;
ATO = average treatment for the overlap population.
\end{table}

Now consider the \textit{incremental propensity score intervention}
(IPSI) proposed in \citet{kennedy2019nonparametric}, shown in the
last row of Table \ref{tab:Four-stylized-interventions}:
\begin{align}
\pi^{\mathcal{I}_{\delta}}(x) & =\frac{e^{\delta}\pi(x)}{1-\pi(x)+e^{\delta}\pi(x)}.\label{eq:IPSI}
\end{align}
Here, we reparameterize the $\delta$ parameter in Equation (1) of
\citet{kennedy2019nonparametric} as $e^{\delta}$, so that $\mathcal{I}_{0}$
corresponds to the status quo. The IPSI is notable for several reasons.
First, Equation \eqref{eq:IPSI} shows that $\pi^{\mathcal{I}_{\delta}}(x)\neq\pi(x)$
only when $\pi(x)\in(0,1)$. Thus, the corresponding IE and MIE are
identified even if the positivity assumption does not hold, because
units with baseline propensity scores of zero or one are not affected
by the intervention. In this regard, the IPSI resembles intervention
(b).

Second, as Kennedy noted, $e^{\delta}$ has a natural interpretation
as an \textit{odds ratio}, describing how the intervention shifts
treatment odds:
\[
e^{\delta}=\frac{\pi^{\mathcal{I}_{\delta}}(x)/\big(1-\pi^{\mathcal{I}_{\delta}}(x)\big)}{\pi(x)/\big(1-\pi(x)\big)}.
\]
That is, the IPSI multiplies everyone's odds of treatment by the same
factor $e^{\delta}$, or equivalently, it increases everyone's \textit{log-odds}
of treatment by $\delta$. This interpretation connects $\pi^{\mathcal{I}_{\delta}}(x)$
to a latent index model in which 
\begin{align*}
A^{\mathcal{I}_{\delta}} & =\mathbb{I}(\delta+g(X)-V\geq0),
\end{align*}
In this model, $g(X)=\textup{logit}\big(\pi(X)\big)$, and $V$ follows
a standard logistic distribution. Under this model, treatment status
$A^{\mathcal{I}_{\delta}}$ is deterministically assigned given $(X,V)$.
This interpretation differs from Kennedy's original interpretation
of the IPSI as a stochastic intervention, where treatment is an independent
Bernoulli draw with probability $\pi^{\mathcal{I}_{\delta}}(X)$.
In contexts where treatment status cannot literally be randomized,
the latent index interpretation may be more plausible.

Third, under the IPSI, 
\begin{align*}
\dot{\pi}^{\mathcal{I}_{0}}(x) & =\pi(x)(1-\pi(x)),
\end{align*}
so the largest increments occur for units with baseline propensity
scores near 0.5. The corresponding MIE becomes
\begin{equation}
\textup{MIE}_{\textup{IPSI}}=\frac{\mathbb{E}[\pi(X)(1-\pi(X))\textup{CATE}(X)]}{\mathbb{E}[\pi(X)(1-\pi(X))]}.\label{eq:ATO}
\end{equation}
This estimand coincides with the \textit{average treatment effect
for the overlap population} (ATO) proposed by \citet{li2018balancing}.
These authors formulate the ATO as the estimand associated with overlap
weights, an alternative to inverse probability weighting in which
``each unit's weight is proportional to the probability of that unit
being assigned to the opposite group'' (p. 390). Moreover, they show
that when $\pi(X)$ is estimated via a logistic regression model,
overlap weights lead to exact balance between treated and untreated
units in the means of all pretreatment covariates. The ATO is akin
to the \textit{optimally weighted average treatment effect} (OWATE;
\citealt{crump2006moving}). Specifically, \citet{crump2006moving}
consider a class of weighted sample average treatment effects in the
form of $\tau_{S,w}=\sum_{i}w(X_{i})\textup{CATE}(X_{i}))/\sum_{i}w(X_{i})$
and show $w^{*}(X)=\pi(X)(1-\pi(X))$ to be the weight that minimizes
$\tau_{S,w}$'s nonparametric variance bound under homoskedasticity
(i.e., $\textup{Var}[Y|X,A]=\textup{constant}$).

Finally, Equation \eqref{eq:ATO} is also often known as the ``regression
estimand'' (\citealt{angrist2008mostly}), defined as the probability
limit of the coefficient on $A$ in a multiple regression of $Y$
on $A$ and $X$ when $\mathbb{E}[A|X]$ is linear in $X$. While
prior work has emphasized its ``nonrepresentative nature'' (e.g.,
\citealt{aronow2016does}), we can see that the regression estimand
is by no means vacuous. Although it is nonrepresentative of the full
population, it corresponds to\textit{ a meaningful, policy-relevant
subpopulation}: those who would be induced into treatment under the
IPSI---an intervention that preserves the relative odds of treatment
across units.

\subsection{Connection with Modified Treatment Policies}

As noted earlier, under our identification assumptions, interventions
that share the same form of $\pi^{\mathcal{I}_{\delta}}(X)$ yield
identical values of $\textup{IE}_{\delta}$ and MIE, which motivated
our use of $\pi^{\mathcal{I}_{\delta}}(X)$ to classify different
types of interventions. Still, even when two interventions induce
the same $\pi^{\mathcal{I}_{\delta}}(X)$, they may be understood
and operationalized in different ways. For example, we saw that the
IPSI can be interpreted as either a deterministic intervention within
a latent index model or a stochastic intervention where each unit's
treatment status is independently drawn from $\textup{Bernoulli}(\pi^{\mathcal{I}_{\delta}}(X))$.
In both cases, the treatment assignment mechanism under $\mathcal{I}_{\delta}$
is assumed to depend only on pretreatment characteristics---$(X,V)$
in the latent index model or $X$ in the stochastic intervention.

In epidemiology, however, scholars have proposed an alternative class
of interventions, known as modified treatment policies (MTPs) (e.g.,
\citealt{robins2004effects,taubman2009intervening,haneuse2013estimation,young2014identification,diaz2023nonparametric,diaz2024causal}).
Under an MTP, treatment status depends, either deterministically or
stochastically, on the ``natural'' treatment value that would have
been observed without the intervention, i.e., $A$. \citet{haneuse2013estimation},
for example, considered the impact of a deterministic MTP that shortens
the operating time by a modest amount, i.e., $A^{\mathcal{I}_{\delta}}=A-\delta$,
on postoperative outcomes among lung cancer patients. The daily exercise
intervention envisioned by \citet{taubman2009intervening} is also
a deterministic MTP, defined as $A^{\mathcal{I}}=\max(30,A)$, although
it is not indexed by a scalar.

At first glance, MTPs appear different from the interventions we have
discussed, since they explicitly condition on the ``natural'' treatment
status $A$. However, our framework for identifying the IE and MIE
does not preclude this possibility. Indeed, the same identification
results apply to MTPs as long as Assumptions \ref{assu:Consistency}-\ref{assu:Support}
hold. In fact, if unconfoundedness holds under the status quo (i.e.,
$Y(a)\ci A|X$), then it also holds under an MTP. This is because
under an MTP, $A^{\mathcal{I}_{\delta}}|X,A$ is either fixed or an
independent random draw, implying $A^{\mathcal{I}_{\delta}}\ci Y(a)|X,A$.
The latter conditional independence, when combined with $Y(a)\ci A|X$,
implies $A^{\mathcal{I}_{\delta}}\ci Y(a)|X$. Thus, the IE and MIE
under an MTP can also be identified using Equations \eqref{eq:IE-CATE}
and \eqref{eq:MIE-CATE}, where the interventional propensity score
$\pi^{\mathcal{I}_{\delta}}(X)$ can be obtained by marginalizing
over $A$:
\[
\pi^{\mathcal{I}_{\delta}}(x)=\pi(x)\Pr[A^{\mathcal{I}_{\delta}}=1|x,A=1]+\big(1-\pi(x)\big)\Pr[A^{\mathcal{I}_{\delta}}=1|x,A=0].
\]
This result echoes \citet{young2014identification}, who show that
the expected outcome under an MTP can be identified using the same
g-formula that one would normally use for a stochastic intervention
that does not depend on $A$, provided that $A$ does not affect the
outcome other than through $A^{\mathcal{I}_{\delta}}$. In our framework,
this condition is guaranteed by system invariance (Assumption \ref{assu:System-invariance}).

To gain more intuition for how MTPs fit into our framework, consider
a stylized MTP where all treated units remain treated, while untreated
units are induced into treatment with probability $\delta$. That
is, $\Pr[A^{\mathcal{I}_{\delta}}=1|x,A=1]=1$ and $\Pr[A^{\mathcal{I}_{\delta}}=1|x,A=0]=\delta$,
which implies $\pi^{\mathcal{I}_{\delta}}(x)=\pi(x)+\delta\big(1-\pi(x)\big)$.
Thus, for this MTP, $\dot{\pi}^{\mathcal{I}_{0}}(x)=1-\pi(x)$, so
the corresponding MIE equals the ATU, equivalent to the third stylized
intervention shown in Table \ref{tab:Four-stylized-interventions}.
By contrast, if untreated units are induced into treatment with probability
proportional to their baseline propensity score, that is, $\Pr[A^{\mathcal{I}_{\delta}}=1|x,A=0]=\delta\pi(x)$,
then $\pi^{\mathcal{I}_{\delta}}(x)=\pi(x)+\delta\pi(x)(1-\pi(x))$.
In this case, $\dot{\pi}^{\mathcal{I}_{0}}(x)=\pi(x)(1-\pi(x))$,
leading to $\textup{MIE}_{\textup{IPSI}}$.

\subsection{Estimation}

For the first three stylized interventions shown in Table \ref{tab:Four-stylized-interventions},
estimation of the MIE is straightforward when positivity holds (or
is assumed to hold). In such cases, we can directly apply existing
estimators of the ATE, ATT, or ATU, such as regression-imputation
(RI; e.g., \citealt{hahn1998role}), inverse probability weighting
(e.g., \citealt{hirano2003efficient}), and doubly robust methods
(\citealt{robins1995semiparametric}).

When positivity is violated but Assumption \ref{assu:Support} holds,
the MIE can still be identified for intervention (b). In this case,
it is equivalent to the ATT conditional on $0<\pi(X)<1$. Such estimands,
however, are non-smooth functionals of the distribution of $(X,A,Y)$,
which complicates estimation and inference. A practical approach in
this setting is to apply a standard ATT estimator to a trimmed sample
that excludes covariate regions where positivity is presumably violated.
For instance, one can first estimate the propensity score and then
restrict the analytic sample to units whose estimated scores lie within
the interval $[\min_{i:A_{i}=1}\hat{\pi}(X_{i}),\max_{i:A_{i}=0}\hat{\pi}(X_{i})]$.
This trimming strategy is widely used in practice to mitigate positivity
violations in conventional causal inference (e.g., \citealt{dehejia1999causal,Carneiro2011}).
In our context, this practice is further justified by the fact that,
when Assumption \ref{assu:Support} holds, MIE weights are zero for
units with propensities scores equal to zero or one. That said, trimming
introduces its own complications. Because the true propensity scores
are unknown, trimming must rely on estimated scores, which can in
turn induce bias and complicate inference. These challenges have been
discussed extensively in the literature (e.g., \citealt{crump2006moving},
\citealt{petersen2012diagnosing}).

For the IPSI, the MIE corresponds to Equation \eqref{eq:ATO}, for
which \citet{li2018balancing} proposed the following weighting estimator:
\[
\textup{\ensuremath{\widehat{\textup{MIE}}}}_{\textup{IPSI}}^{\textup{weighting}}\textup{=}\frac{\sum_{i}(1-\hat{\pi}(X_{i}))A_{i}Y_{i}}{\sum_{i}(1-\hat{\pi}(X_{i}))A_{i}}-\frac{\sum_{i}\hat{\pi}(X_{i})(1-A_{i})Y_{i}}{\sum_{i}\hat{\pi}(X_{i})(1-A_{i})}.
\]
An appealing feature of this estimator is that, when the propensity
score $\pi(X)$ is estimated via a logistic regression model, it achieves
exact balance between treated and untreated units in the means of
the covariates. Alternatively, one can use an RI approach described
in \citet{crump2006moving}: 
\[
\textup{\ensuremath{\widehat{\textup{MIE}}}}_{\textup{IPSI}}^{\textup{RI}}\textup{=}\frac{\sum_{i}\hat{\pi}(X_{i})(1-\hat{\pi}(X_{i}))(\hat{\mu}_{1}(X_{i})-\hat{\mu}_{0}(X_{i}))}{\sum_{i}\hat{\pi}(X_{i})(1-\hat{\pi}(X_{i}))},
\]
where $\mu_{a}(X)\stackrel{\Delta}{=}\mathbb{E}[Y|X,A=a]$ for $a=0,1$.
For both the weighting and RI estimators of $\textup{MIE}_{\textup{IPSI}}$,
standard errors and confidence intervals can be obtained using the
nonparametric bootstrap. In addition, \citet{crump2006moving} show
that the RI estimator achieves the nonparametric efficiency bound
if both the propensity score and outcome models are estimated nonparametrically
(e.g., with kernel methods) and certain technical conditions hold.
In that case, analytical standard errors can also be derived.

Besides the weighting and RI estimators, Equation \eqref{eq:ATO}
can also be estimated using Robinson's (\citeyear{robinson1988root})
``partialing-out'' estimator :
\begin{equation}
\textup{\ensuremath{\widehat{\textup{MIE}}}}_{\textup{IPSI}}^{\textup{Robinson}}\textup{=}\frac{\sum_{i}(Y_{i}-\hat{\mathbb{E}}[Y_{i}|X_{i}])(A_{i}-\hat{\mathbb{E}}[A_{i}|X_{i}])}{\sum_{i}(A_{i}-\hat{\mathbb{E}}[A_{i}|X_{i}])^{2}}.\label{eq:robinson}
\end{equation}
The partialing-out estimator arises in the context of the partially
linear regression (PLR) model. Although the PLR assumes a constant
treatment effect, the probability limit of the partialing-out estimator
remains a well-defined statistical parameter under treatment effect
heterogeneity. Specifically, it is equal to 
\[
\tau^{\textup{Robinson}}=\frac{\mathbb{E}[\textup{Cov}[Y,A|X]]}{\mathbb{E}[\textup{Var}[A|X]]},
\]
which, when $A$ is binary, reduces to Equation \eqref{eq:ATO} (\citealt{vansteelandt2020assumption}).

An advantage of $\textup{\ensuremath{\widehat{\textup{MIE}}}}_{\textup{IPSI}}^{\textup{Robinson}}$
over the weighting and RI estimators is that its estimating equation
\eqref{eq:robinson} is based on the efficient influence function
(EIF) of $\tau^{\textup{Robinson}}$ in the nonparametric model:
\[
\frac{(A-\mathbb{E}[A|X])\big(Y-\mathbb{E}[Y|X]-\tau^{\textup{Robinson}}(A-\mathbb{E}[A|X])\big)}{\textup{Var}[A|X]}.
\]
Consequently, $\textup{\ensuremath{\widehat{\textup{MIE}}}}_{\textup{IPSI}}^{\textup{Robinson}}$
is ``Neyman-orthogonal'' (\citealt{chernozhukov2018double}), meaning
that first step estimation of the nuisance functions $\hat{\mathbb{E}}[Y|X]$
and $\hat{\mathbb{E}}[A|X]$ has no first-order effect on the influence
function of the estimator. This property allows $\textup{\ensuremath{\widehat{\textup{MIE}}}}_{\textup{IPSI}}^{\textup{Robinson}}$
to achieve $\sqrt{n}$-consistency even when the nuisance functions
are estimated using flexible, data-adaptive methods, facilitating
what Chernozhukov et al. call debiased machine learning (DML). In
particular, if both nuisance function estimates are consistent and
converge at faster-than-$n^{-1/4}$ rates, and if cross-fitting is
used to render the empirical process term asymptotically negligible,
then $\textup{\ensuremath{\widehat{\textup{MIE}}}}_{\textup{IPSI}}^{\textup{Robinson}}$
is $\sqrt{n}$-consistent, asymptotically normal, and semiparametrically
efficient. Standard errors can then be estimated via the empirical
variance of its estimated EIF.

For all four stylized interventions described above, the interventional
propensity score $\pi^{\mathcal{I}_{\delta}}(X)$ is a known function
of the baseline propensity score; so is its local derivative $\dot{\pi}^{\mathcal{I}_{0}}(X)=\partial\pi^{\mathcal{I}_{\delta}}(X)/\partial\delta|_{\delta=0}$.
In general, if $\dot{\pi}^{\mathcal{I}_{0}}(X)=\lambda(\pi(X))$ for
a known function $\lambda(\cdot)$, the MIE (i.e., Equation \ref{eq:MIE-CATE})
can be estimated by the following RI estimator:
\begin{equation}
\textup{\ensuremath{\widehat{\textup{MIE}}}}^{\textup{RI}}\textup{=}\frac{\sum_{i}\lambda(\hat{\pi}(X_{i}))(\hat{\mu}_{1}(X_{i})-\hat{\mu}_{0}(X_{i}))}{\sum_{i}\lambda(\hat{\pi}(X_{i}))}.\label{eq:MIE-RI}
\end{equation}
Alternatively, one can construct a Neyman-orthogonal estimating equation
based on the efficient influence function (EIF) of the corresponding
MIE (see Crump et al. 2006), allowing the nuisance functions to be
estimated using flexible, data-adaptive methods.

Finally, for the special case where $\dot{\pi}^{\mathcal{I}_{0}}(X)$
itself is known---such as in a randomized experiment where $\pi^{\mathcal{I}_{\delta}}(X)$
is fully specified---one can either use an RI estimator similar to
Equation \eqref{eq:MIE-RI} (replacing $\lambda(\hat{\pi}(X_{i}))$
with $\dot{\pi}^{\mathcal{I}_{0}}(X_{i})$) or construct a Neyman-orthogonal
and doubly robust estimating equation based on its EIF (see \citealt{hirano2003efficient}).

\subsection{Illustration}

We illustrate the MIE estimators described above using data from a
right heart catheterization (RHC) study in the U.S. (Connors et al.,
1996), which collected survival outcomes for 5,735 hospitalized adult
patients. RHC is a diagnostic procedure that directly measures cardiac
function in critically ill patients but carries a risk of clinical
complications. Although treatment assignment was not randomized, Connors
et al. (1996) used propensity-score-based matching and multivariate
regression modeling to adjust for confounding and found that RHC was
associated with an increase in 30-day mortality.

We use all pretreatment covariates $X$ potentially related to the
decision to use RHC, as recorded in the study (full variable descriptions
are available at \url{ https://biostat.app.vumc.org/wiki/pub/Main/DataSets/rhc.html}.
Among the 65 covariates (21 continuous, 24 binary, and 20 dummies
derived from 5 categorical variables), several key covariates show
substantial differences in distribution between the treated and untreated
units (\citealt{hirano2001estimation}, Table 2). Our outcome of interest
is survival 180 days after potential receipt of RHC. This dataset
was most recently analyzed by \citet{li2018balancing}, who applied
weighting estimators to estimate the ATE, ATT, and ATO.

We conduct two sets of analyses. First, we assume that positivity
holds, i.e., $\textup{supp}(\pi(X))\subset(0,1)$. Under this assumption,
the MIEs for the four stylized interventions coincide with the ATE,
ATT, ATU, and ATO, and all are nonparametrically identified. Second,
we relax the positivity assumption so that some units may have propensity
scores of exactly zero or one. In this setting, the ATE, ATT, and
ATU are no longer identified. However, the MIE remains identified
under the second and fourth stylized interventions---namely, when
$\pi^{\mathcal{I}_{\delta}}(X)=\min\{1,\pi(x)e^{\delta}\}$ and $\pi^{\mathcal{I}_{\delta}}(X)=\frac{e^{\delta}\pi(x)}{1-\pi(x)+e^{\delta}\pi(x)}$.
When positivity is violated, units with propensity scores of zero
or one must be excluded. However, because the true propensity scores
are unknown, we operationalize this by trimming units whose estimated
propensity scores fall below $\min_{i:A_{i}=1}\hat{\pi}(X_{i})$ or
above $\textup{max}_{i:A_{i}=0}\hat{\pi}(X_{i})$ , on the presumption
that these units would violate positivity. The estimated propensity
scores $\hat{\pi}(X_{i})$ are obtained from a logistic regression
of treatment on all covariates. In our data, $\min_{i:A_{i}=1}\hat{\pi}(X_{i})=0.019$
and $\textup{max}_{i:A_{i}=0}\hat{\pi}(X_{i})=0.961$, leading us
to exclude 65 untreated units and 5 treated units.

For each MIE, we apply three estimators. First, we employ a parametric
IPW approach, using standard ATE, ATT, and ATU weights for the first
three interventions and the ATO weight (\citealt{li2018balancing})
for the IPSI. In all cases, we estimate the propensity scores using
a logistic regression of treatment on all covariates. Next, we use
a parametric RI approach, where the outcome model is estimated using
a logistic regression that includes the treatment and all covariates.
Note that this specification of the outcome model implies a constant
treatment effect on the log-odds scale but not on the probability
scale. Finally, we implement a DML procedure for each of the four
interventions. Specifically, for the first three interventions, we
use the doubly robust estimating equations for the ATE, ATT, and ATU
described in \citet{chernozhukov2018double}. For the IPSI, we use
Robinson's partialing-out estimator \eqref{eq:robinson}. In the DML
approach, the outcome and propensity score models are both fit using
a super learner (\citealt{van2007super}) composed of the generalized
linear model (GLM), Lasso, and random forest, and the final estimates
are obtained using five-fold cross-fitting. Standard errors for the
IPW and RI estimators are estimated using the nonparametric bootstrap
with 1,000 replications; standard errors for the DML estimators are
estimated from the sample variances of the estimated EIFs.

Table \ref{tab:Results1} presents our estimates of the MIE for the
four stylized interventions. To make our results comparable to those
reported in \citet{li2018balancing}, we multiply our point estimates
and standard errors (which are on the probability scale) by $100$.
All of our estimators suggest that receiving RHC leads to a higher
mortality rate than not applying RHC. We find that the MIE estimates
under DML are all somewhat smaller than those obtained from the parametric
IPW and RI approaches; moreover, despite the use of machine learning
methods to reduce model misspecification bias, the estimated standard
errors under DML are comparable to those under parametric RI and,
except for the ATO, considerably smaller than those under parametric
IPW. Under both parametric IPW and DML, our point estimate for the
ATO is higher than that for the ATE (in absolute value), suggesting
that an IPSI may induce into treatment individuals who are particularly
vulnerable to its harmful effects. Finally, we can see that across
all estimators, standard errors tend to be smallest when the estimand
is the ATO, consistent with its optimality property established in
previous work (\citealt{crump2006moving,li2018balancing}). 

\begin{table}
\caption{MIE estimates under four stylized interventions for the RHC data.\label{tab:Results1}}

\begin{centering}
{\small{}%
\begin{tabular}{l>{\centering}p{1.6cm}>{\centering}p{1.6cm}>{\centering}p{1.6cm}>{\centering}p{1.6cm}>{\centering}p{1.6cm}>{\centering}p{1.6cm}}
\hline 
\multirow{2}{*}{{\small$\pi^{\mathcal{I}_{\delta}}(x)$}} & \multicolumn{2}{c}{parametric IPW} & \multicolumn{2}{c}{parametric RI} & \multicolumn{2}{c}{DML}\tabularnewline
\cline{2-7}
 & assuming positivity & relaxing positivity & assuming positivity & relaxing positivity & assuming positivity & relaxing positivity\tabularnewline
\hline 
{\small$\min\{1,\pi(x)+\delta\}$} & -5.43 (1.68) &  & -5.72 (1.33) &  & -4.64 (1.20) & \tabularnewline
{\small$\min\{1,\pi(x)e^{\delta}\}$} & -5.62 (1.89) & -4.89 (1.81) & -5.80 (1.37) & -5.93 (1.34) & -5.23 (1.30) & -4.78 (1.29)\tabularnewline
{\small$1-(1-\pi(x))e^{-\delta}$} & -5.32 (2.13) &  & -5.67 (1.32) &  & -4.27 (1.32) & \tabularnewline
{\small$\frac{e^{\delta}\pi(x)}{1-\pi(x)+e^{\delta}\pi(x)}$} & -5.88 (1.35) & -5.88 (1.32) & -5.76 (1.32) & -5.88 (1.29) & -5.12 (1.32) & -4.89 (1.32)\tabularnewline
\hline 
\end{tabular}}{\small\smallskip{}
}{\small\par}
\par\end{centering}
Note: MIE = marginal interventional effect; RHC = right heart catheterization;
IPW = inverse probability weighting; RI = regression-imputation; DML
= debiased machine learning. Numbers in parentheses are standard errors.
\end{table}

Compare results from the full sample (assuming positivity) and the
trimmed sample (relaxing positivity), we find that trimming has little
effect on the substantive conclusions. Under both parametric IPW and
DML, the MIE estimates for the multiplicative propensity score intervention
($\pi^{\mathcal{I}_{\delta}}(X)=\min\{1,\pi(x)e^{\delta}\}$) are
slightly attenuated in absolute value when positivity is relaxed,
while those under the IPSI ($\pi^{\mathcal{I}_{\delta}}(X)=\frac{e^{\delta}\pi(x)}{1-\pi(x)+e^{\delta}\pi(x)}$)
remain largely unchanged. Thus, in this example, excluding units with
extreme estimated propensity scores reduces reliance on extrapolation
without materially altering the overall message: RHC is associated
with increased mortality risk, and this adverse effect may be especially
pronounced under interventions that target the overlap population.

\section{Identification and Estimation with Instrumental Variables}

\label{sec:IV}

\noindent From Section \ref{sec:Unconfoundedness}, we have seen that
under unconfoundedness, the IE and MIE can be identified as weighted
means of CATEs. The unconfoundedness assumption, however, is strong,
untestable, and implausible in many applications. In observational
studies where unconfoundedness is doubtful given subject matter knowledge,
or in randomized trials with treatment noncompliance, researchers
often seek to identify causal effects using an instrumental-variable
(IV) approach (which entails an alternative set of strong and untestable
assumptions). In what follows, we demonstrate that the IE and MIE
can also be identified with IVs within the framework of marginal treatment
effects (MTE; \citealt{HeckmanVytlacil1999,HeckmanVytlacil2005}).

\subsection{The Generalized Roy Model}

The MTE framework builds on the generalized Roy model for discrete
choices (\citealt{roy1951some,HeckmanVytlacil2005}). As before, let
$A$ denote a binary treatment, $Y(a)$ the potential outcome under
treatment status $a$, and $X$ a vector of pretreatment covariates.
Following \citet{zhou2019marginal}, we write the potential outcome
equations as
\begin{align}
Y(0) & =\mu_{0}(X)+\epsilon,\label{eq:Y0}\\
Y(1) & =\mu_{1}(X)+\epsilon+\eta,\label{eq:Y1}
\end{align}
where $\mu_{0}(X)=\mathbb{E}[Y(0)|X]$, $\mu_{1}(X)=\mathbb{E}[Y(1)|X]$,
the error term $\epsilon$ captures unobserved factors affecting the
baseline outcome ($Y(0)$), and the error term $\eta$ captures all
unobserved factors affecting the treatment effect ($Y(1)-Y(0)$).
In general, the error terms $\epsilon$ and $\eta$ need not be statistically
independent of $X$, although they have zero conditional means by
construction. Under Assumption \ref{assu:Consistency} (\textit{consistency}),
the observed outcome $Y$ can be expressed through a switching regression
model (\citealt{quandt1958estimation,quandt1972new}):
\begin{align}
Y & =(1-A)Y(0)+AY(1)\nonumber \\
 & =\mu_{0}(X)+(\mu_{1}(X)-\mu_{0}(X))A+\epsilon+\eta A.\label{eq:Yobs}
\end{align}

Treatment assignment is represented by a latent index model. Let $I_{A}$
denote a latent tendency for treatment, which depends on both observed
($Z$) and unobserved ($V$) factors:
\begin{align}
I_{A} & =\nu(Z)-V,\label{eq:IA}\\
A & =\mathbb{I}(I_{A}\geq0),\label{eq:A}
\end{align}
where $\nu(Z)$ is an unspecified function, $V$ is a continuous latent
random variable with a strictly increasing distribution function,
representing unobserved resistance to treatment. The $Z$ vector includes
all components of $X$ as well as excluded instruments ($Z\backslash X$),
i.e., variables that affect treatment assignment but not potential
outcomes. Moreover, we assume that at least one component of $Z\backslash X$
is continuous. A continuous IV is needed because it generates smooth
variation in the propensity score $\pi(Z)=\Pr[A=1|Z]$, which in turn
allows us to identify the MTE as a partial derivative of $\mathbb{E}[Y|X,Z]$
with respect to $\pi(Z)$ (see Section 4.2). 

The key assumptions associated with Equations \eqref{eq:Y0}-\eqref{eq:A}
are
\begin{assumption}
Independence: $(\epsilon,\eta,V)\ci Z|X$.\label{assu:Independence}
\end{assumption}
\begin{assumption}
Relevance: $\nu(Z)$ is a nondegenerate function of $Z$ given $X$.\label{assu:Rank-condition}
\end{assumption}
\noindent The latent index model characterized by Equations \eqref{eq:IA}
and \eqref{eq:A}, combined with Assumptions \ref{assu:Independence}-\ref{assu:Rank-condition},
is equivalent to the Imbens-Angrist \citeyearpar{ImbensAngrist1994}
assumptions of independence and monotonicity for interpreting IV estimands
as local average treatment effects (LATE) (\citealt{vytlacil2002independence}).
In particular, the separability of $\nu(Z)$ and $V$ in Equation
\eqref{eq:IA} implies that a change in $Z$ (say, from $z_{1}$ to
$z_{2}$) shifts the latent tendency $I_{A}$ in the same direction
for all units, thereby ruling out the presence of ``defiers.'' Moreover,
under Assumptions \ref{assu:Independence}-\ref{assu:Rank-condition},
the latent resistance $V$ is allowed to be correlated with $\epsilon$
and $\eta$ in a general way. For example, in research on heterogeneous
returns to schooling, it has been argued that individuals may self-select
into college based on anticipated gains. In this setting, $V$ would
be negatively correlated with $\eta$, since individuals with higher
values of $\eta$ (greater potential returns) tend to have lower unobserved
resistance to schooling.

\subsection{Marginal Treatment Effects}

To define the MTE, we rewrite the treatment assignment equations \eqref{eq:IA}
and \eqref{eq:A} as 
\begin{align}
A & =\mathbb{I}(F_{V|X}(\nu(Z))-F_{V|X}(V)\geq0)\nonumber \\
 & =\mathbb{I}(\pi(Z)-U\geq0),\label{eq:p-u}
\end{align}
where $F_{V|X}(\cdot)$ is the cumulative distribution function (CDF)
of $V$ given $X$, and $\pi(Z)=\textup{Pr}(A=1|Z)=F_{V|X}(\nu(Z))$
denotes the propensity score given $Z$. By definition, $U=F_{V|X}(V)$
follows a standard uniform distribution. Equation \eqref{eq:p-u}
shows that $Z$ affects treatment status only through the propensity
score $\pi(Z)$. This property---that instruments influence treatment
solely through an additively separable index---is known as index
sufficiency (\citealt{HeckmanVytlacil2005}).

The MTE is defined as the expected treatment effect conditional on
pretreatment covariates $X=x$ and the ``normalized'' latent variable
$U=u$:
\begin{align}
\textup{MTE}(x,u) & =\mathbb{E}[Y(1)-Y(0)|X=x,U=u]\nonumber \\
 & =\mathbb{E}[\mu_{1}(X)-\mu_{0}(X)+\eta|X=x,U=u]\nonumber \\
 & =\mu_{1}(x)-\mu_{0}(x)+\mathbb{E}[\eta|X=x,U=u].\label{eq:mte}
\end{align}
Because $U$ is the CDF of $V$, variation in $\textup{MTE}(x,u)$
across values of $u$ reflects how treatment effects differ across
quantiles of the unobserved resistance to treatment, conditional on
$X$. Equivalently, $\textup{MTE}(x,u)$ can be interpreted as the
average treatment effect among individuals with covariates $X=x$
and the propensity score $\pi(Z)=u$ who are exactly indifferent between
taking treatment or not (i.e., those for whom $U=\pi(Z)$). 

Conventional causal estimands, such as the ATE and ATT, can be expressed
as weighted averages of $\textup{MTE}(x,u)$. To obtain population-level
effects, $\textup{MTE}(x,u)$ needs to be marginalized twice, first
over the distribution of $U$ given $X$ and then over the marginal
distribution of $X$. The weighting functions that link the MTE to
the ATE, ATT, and ATU are provided in \citet{HeckmanUrzuaVytlacil2006a}.

Given Equations \eqref{eq:Y0}-\eqref{eq:A} and Assumptions \ref{assu:Consistency},
\ref{assu:Independence}, and \ref{assu:Rank-condition}, $\textup{MTE}(x,u)$
can be identified using the method of local instrumental variables
(LIV). To see this, let us consider the conditional mean of the observed
outcome $Y$ given $X=x$ and the propensity score $\pi(Z)=p$. According
to Equation \eqref{eq:Yobs}, we have
\begin{align}
\mathbb{E}[Y|X=x,\pi(Z)=p] & =\mathbb{E}[\mu_{0}(X)+(\mu_{1}(X)-\mu_{0}(X))A+\epsilon+\eta A|X=x,\pi(Z)=p]\nonumber \\
 & =\mu_{0}(x)+(\mu_{1}(x)-\mu_{0}(x))p+\mathbb{E}[\eta|A=1,X=x,\pi(Z)=p]p\nonumber \\
 & =\mu_{0}(x)+(\mu_{1}(x)-\mu_{0}(x))p+\int_{0}^{p}\mathbb{E}[\eta|X=x,U=u]du\nonumber \\
 & =\mu_{0}(x)+\int_{0}^{p}\textup{MTE}(x,u)du.\label{eq:Yxp}
\end{align}
Thus, the MTE can be identified as the partial derivative of this
conditional expectation with respect to $p$:
\begin{align}
\textup{MTE}(x,p) & =\frac{\partial\mathbb{E}[Y|X=x,\pi(Z)=p]}{\partial p}.\label{eq:localIV}
\end{align}
Since $\mathbb{E}[Y|X=x,\pi(Z)=p]$ is a function of observed data,
this expression implies that $\textup{MTE}(x,u)$ is identified at
all values of $u$ within the conditional support $\textup{supp}(\pi(Z)|X=x)$.
Equivalently, $\textup{MTE}(x,u)$ is identified over $\textup{supp}(X,\pi(Z))$,
the support of the joint distribution of $X$ and $\pi(Z)$. Crucially,
this differentiation step requires that at least one IV be continuously
distributed, so that $\pi(Z)$ varies smoothly with $Z$.

The MTE also provides a unifying framework for interpreting the local
average treatment effect (LATE) of Imbens and Angrist (1994). Specifically,
the LATE for compliers induced by a change in the instrument from
$z_{1}$ to $z_{2}$ can be written as (\citealt{HeckmanVytlacil2005}):
\[
\text{LATE}(x;z_{1},z_{2})=\frac{1}{\pi(z_{2})-\pi(z_{1})}\int_{\pi(z_{1})}^{\pi(z_{2})}\text{MTE}(x,u)\,du.
\]
That is, the LATE corresponds to a weighted average of $\textup{MTE}(x,u)$
over an interval of $u$, corresponding to individuals whose treatment
status changes when the instrument moves from $z_{1}$ to $z_{2}$.
This formulation makes clear that the LATE is a special case of the
MTE framework: while the LATE recovers the average treatment effect
for compliers under discrete instrument shifts, the MTE traces out
the full distribution of treatment effects across quantiles of unobserved
resistance, provided that $\pi(Z)$ varies continuously. 

\subsection{Identification of IE and MIE}

As in Section \ref{sec:Unconfoundedness}, we focus on interventions
that satisfy Assumptions \ref{assu:Consistency}-\ref{assu:Differentiability}.
We again assume that $\dot{\pi}^{\mathcal{I}_{0}}(z)=\partial\pi^{\mathcal{I}_{\delta}}(z)/\partial\delta|_{\delta=0}$
exists for all $z\in\textup{supp}(Z)$, and that it is either a known
function of $z$ directly or a known function of $x$ and the baseline
propensity score $\pi(z)$. 

Assumptions \ref{assu:Consistency}-\ref{assu:System-invariance}
imply that the outcome models \eqref{eq:Y0}-\eqref{eq:Y1} are invariant
under interventions. Thus the observed outcome $Y^{\mathcal{I}_{\delta}}$
under intervention $\mathcal{I}_{\delta}$ can be written as
\begin{align}
Y^{\mathcal{I}_{\delta}} & =(1-A^{\mathcal{I}_{\delta}})Y(0)+A^{\mathcal{I}_{\delta}}Y(1)\nonumber \\
 & =\mu_{0}(X)+(\mu_{1}(X)-\mu_{0}(X))A^{\mathcal{I}_{\delta}}+\epsilon+\eta A^{\mathcal{I}_{\delta}},\label{eq:Ydelta}
\end{align}
Analogous to Equation \eqref{eq:Yxp}, we have
\begin{align*}
\mathbb{E}[Y^{\mathcal{I}_{\delta}}] & =\mathbb{E}\mathbb{E}[\mu_{0}(X)+(\mu_{1}(X)-\mu_{0}(X))A^{\mathcal{I}_{\delta}}+\epsilon+\eta A^{\mathcal{I}_{\delta}}|Z]\\
 & =\mathbb{E}\big[\mu_{0}(X)+(\mu_{1}(X)-\mu_{0}(X))\pi^{\mathcal{I}_{\delta}}(Z)+\int_{0}^{\pi^{\mathcal{I}_{\delta}}(Z)}\mathbb{E}[\eta|X,U=u]du\big]\\
 & =\mathbb{E}\big[\mu_{0}(X)+\int_{0}^{\pi^{\mathcal{I}_{\delta}}(Z)}\textup{MTE}\big(X,u\big)du\big].
\end{align*}
Substituting the above expression into Equation \eqref{eq:IE} yields
\begin{align}
\textup{IE}_{\delta} & =\frac{\mathbb{E}[Y^{\mathcal{I}_{\delta}}]-\mathbb{E}[Y^{\mathcal{I}_{0}}]}{\mathbb{E}[A^{\mathcal{I}_{\delta}}]-\mathbb{E}[A^{\mathcal{I}_{0}}]}\nonumber \\
 & =\frac{\mathbb{E}\big[\int_{\pi(Z)}^{\pi^{\mathcal{I}_{\delta}}(Z)}\textup{MTE}(X,u)du\big]}{\mathbb{E}\big[\pi^{\mathcal{I}_{\delta}}(Z)-\pi(Z)\big]}.\label{eq:IE-MTE}
\end{align}
Thus, $\textup{IE}_{\delta}$ is identified if, for each unit, $\textup{MTE}(X,u)$
is identified for all $u\in[\pi(Z),\pi^{\mathcal{I}_{\delta}}(Z)]$.
Because $\textup{MTE}(x,u)$ is identified for all $u\in\textup{supp}(\pi(Z)|X=x)$,
a sufficient condition for $\textup{IE}_{\delta}$ to be identified
is therefore 
\begin{align*}
\textup{supp}(\pi^{\mathcal{I}_{\delta}}(Z)|X & =x)\subset\textup{supp}(\pi(Z)|X=x),
\end{align*}
that is, given each covariate value $x$, the support of the interventional
propensity score $\pi^{\mathcal{I}_{\delta}}(Z)$ must lie within
the support of the baseline propensity score $\pi(Z)$.

Intuitively, the above condition requires that the intervention does
not move units outside the region of the propensity score where the
MTE is already identified. It is more likely to be satisfied when
the instruments are strong, i.e., when they generate substantial exogenous
variation in the baseline propensity score conditional on $X$. When
an excluded instrument $Z\backslash X$ has a large and systematic
impact on treatment assignment, the range of $\pi(Z)|X$ is wide,
making it more likely that the interventional propensity score $\pi^{\mathcal{I}_{\delta}}(Z)$
remains within the baseline support. In contrast, weak instruments
produce only limited variation in $\pi(Z)|X$, so interventions may
easily move the propensity score outside the baseline support, violating
the above condition. Put differently, strong instruments ensure that
the support of $\pi(Z)$ given $X$ is rich enough to accommodate
counterfactual policy shifts and thereby sustain identification of
$\textup{IE}_{\delta}.$

In the MTE framework, the IE bears some similarity to the LATE. Both
estimands capture causal effects for individuals whose treatment status
is shifted by a change in the assignment mechanism. For the LATE,
these are the \textquotedblleft compliers\textquotedblright{} whose
treatment is induced by moving the instrument from one value to another.
By contrast, the IE is defined with respect to a counterfactual intervention
that perturbs the propensity score for all units, not just for those
at two discrete instrument values. In this sense, the IE can be viewed
as an analogue of the LATE under policy interventions: both are complier-type
parameters, but whereas the LATE is tied to exogenous variation in
an observed instrument, the IE is tied to a hypothetical intervention
on the assignment mechanism, which may or may not coincide with variation
in the observed instrument.

Under appropriate regularity conditions that allow us to apply L'H\^{o}pital's
rule, the Leibniz integral rule, and exchange differentiation/limits
and integration, the corresponding MIE can be written as
\begin{align}
\textup{MIE} & =\lim_{\delta\downarrow0}\frac{\mathbb{E}[Y^{\mathcal{I}_{\delta}}]-\mathbb{E}[Y^{\mathcal{I}_{0}}]}{\mathbb{E}[A^{\mathcal{I}_{\delta}}]-\mathbb{E}[A^{\mathcal{I}_{0}}]}\nonumber \\
 & =\lim_{\delta\downarrow0}\frac{\frac{\partial\mathbb{E}\big[\int_{\pi(Z)}^{\pi^{\mathcal{I}_{\delta}}(Z)}\textup{MTE}(X,u)du\big]}{\partial\delta}}{\frac{\partial\mathbb{E}\big[\pi^{\mathcal{I}_{\delta}}(Z)-\pi(Z)\big]}{\partial\delta}}\nonumber \\
 & =\lim_{\delta\downarrow0}\frac{\mathbb{E}\big[\frac{\partial\int_{\pi(Z)}^{\pi^{\mathcal{I}_{\delta}}(Z)}\textup{MTE}(X,u)du}{\partial\delta}\big]}{\mathbb{E}\big[\frac{\partial\big[\pi^{\mathcal{I}_{\delta}}(Z)-\pi(Z)\big]}{\partial\delta}\big]}\nonumber \\
 & =\lim_{\delta\downarrow0}\frac{\mathbb{E}\big[\dot{\pi}^{\mathcal{I}_{\delta}}(Z)\textup{MTE}\big(X,\pi^{\mathcal{I}_{\delta}}(Z)\big)\big]}{\mathbb{E}\big[\dot{\pi}^{\mathcal{I}_{\delta}}(Z)\big]}\nonumber \\
 & =\mathbb{E}\big[\frac{\dot{\pi}^{\mathcal{I}_{0}}(Z)}{\mathbb{E}\big[\dot{\pi}^{\mathcal{I}_{0}}(Z)\big]}\textup{MTE}\big(X,\pi(Z)\big)\big].\label{eq:MIE-MTE}
\end{align}
Thus, the MIE is a weighted mean of $\textup{MTE}\big(X,\pi(Z)\big)$.
Similar to the identification formula for the MIE under unconfoundedness,
the weight is proportional to the local derivative of the propensity
score, which gauges the relative intensity of an infinitesimal intervention
among units with $Z=z$. Yet, importantly, the ``building block''
here is not $\textup{CATE}(x)$, but rather $\textup{MTE}\big(x,\pi(z)\big)$---the
marginal treatment effect evaluated at $x$ and $u=\pi(z)$.

A notable implication of Equation \eqref{eq:MIE-MTE} is that, unlike
conventional causal estimands such as the ATE, the MIE is a weighted
mean of $\textup{MTE}\big(X,\pi(Z)\big)$, not of $\textup{MTE}(X,U)$.
This distinction is crucial, as it means support conditions are not
required to identify the MIE, because $\textup{MTE}\big(x,\pi(z)\big)$
can be recovered directly as $\partial\mathbb{E}[Y|x,\pi(z)]/\partial\pi(z)$
over the support of ($X,\pi(Z)$).

Similar to the setting of unconfoundedness, we can envision stylized
interventions where $\dot{\pi}^{\mathcal{I}_{0}}(z)$ is either a
known function of $z$ or a known function of the baseline propensity
score $\pi(z)$. For example, in a ``neutral'' intervention where
$\pi^{\mathcal{I}_{\delta}}(z)=\min\{1,\pi(z)+\delta\}$, $\dot{\pi}^{\mathcal{I}_{0}}(z)$
is equal to $\mathbb{I}(\pi(z)<1)$, and the corresponding MIE will
be an unweighted mean of $\textup{MTE}\big(X,\pi(Z)\big)$ over the
subset of units with $\pi(Z)<1$. Yet, unlike the setting of unconfoundedness,
the MIE under such a neutral intervention does not reduce to the ATE
even if positivity holds. Instead, it reflects the average treatment
effect among units who are at the margin of receiving treatment, i.e.,
$\mathbb{E}[Y(1)-Y(0)|U=\pi(Z)]$.

We can also consider disequalizing or equalizing interventions similar
to those discussed in Section \ref{subsec:Special-Cases}, in which
increments to the propensity score vary systematically with the baseline
propensity. For instance, in an equalizing intervention, individuals
with lower baseline propensity scores receive larger increments, while
those with higher scores receive smaller increments (or none). A real-world
example is a college outreach program targeted toward low-income and
racial minority neighborhoods, which would induce more \textquotedblleft unlikely
college-goers\textquotedblright{} into higher education. Such a policy
would be approximated by an equalizing intervention in which $\dot{\pi}^{\mathcal{I}_{0}}(z)$
is a decreasing function of $\pi(z)$.

Finally, if there is no unobserved selection, we have $\big(Y(1),Y(0)\big)\ci A|X$,
which implies $(\epsilon,\eta)\ci U|X$. In this case, it is clear
that for all $u$,
\begin{align*}
\textup{MTE}(x,u) & =\mathbb{E}[Y(1)-Y(0)|X=x,U=u]\\
 & =\mathbb{E}[Y(1)-Y(0)|X=x]\\
 & =\textup{CATE}(x).
\end{align*}
Substituting this into Equation \eqref{eq:MIE-MTE} yields
\begin{align*}
\textup{MIE} & =\frac{\mathbb{E}\big[\dot{\pi}^{\mathcal{I}_{0}}(Z)\textup{CATE}\big(X\big)\big]}{\mathbb{E}\big[\dot{\pi}^{\mathcal{I}_{0}}(Z)\big]}\\
 & =\frac{\mathbb{E}\big[\mathbb{E}\big[\dot{\pi}^{\mathcal{I}_{0}}(Z)|X\big]\textup{CATE}\big(X\big)\big]}{\mathbb{E}\mathbb{E}\big[\dot{\pi}^{\mathcal{I}_{0}}(Z)|X\big]}\\
 & =\mathbb{E}\big[\frac{\dot{\pi}^{\mathcal{I}_{0}}(X)}{\mathbb{E}[\dot{\pi}^{\mathcal{I}_{0}}(X)]}\textup{CATE}(X)\big],
\end{align*}
where the last equality follows from the fact that $\mathbb{E}\big[\dot{\pi}^{\mathcal{I}_{0}}(Z)|X\big]=\mathbb{E}[\partial\pi^{\mathcal{I}_{\delta}}(Z)/\partial\delta|_{\delta=0}|X]=\partial_{\delta}\mathbb{E}[\pi^{\mathcal{I}_{\delta}}(Z)|X]/\partial\delta|_{\delta=0}=\partial_{\delta}\pi^{\mathcal{I}_{\delta}}(X)/\partial\delta|_{\delta=0}=\dot{\pi}^{\mathcal{I}_{0}}(X)$.
Thus, in the absence of unobserved selection, the identification formula
for MIE reduces to that derived under unconfoundedness, which does
not involve any instrumental variables. A parallel simplification
holds for the IE as well.

\subsection{Estimation}

\label{subsec:Estimation-IV}

\noindent Since we assume that $\dot{\pi}^{\mathcal{I}_{0}}(z)$ is
either a known function of $z$ directly or a known function of $x$
and the baseline propensity score $\pi(z)$, the weights in Equation
\eqref{eq:MIE-MTE} are either known or can be estimated through a
propensity score model. To simplify exposition, we henceforth treat
$\dot{\pi}^{\mathcal{I}_{0}}(z)$ as known, while noting that in practice
it can be replaced with a suitable estimator when unknown. We outline
two approaches to estimating the MIE: a plug-in approach based directly
on Equation \eqref{eq:MIE-MTE} and a doubly robust approach that
applies when $\dot{\pi}^{\mathcal{I}_{0}}(z)$ is a known function
of $x$ and $\pi(z)$.

\subsubsection{Plug-in Approach}

First, Equation \eqref{eq:MIE-MTE} suggests a plug-in estimator of
the MIE:
\begin{align}
\widehat{\textup{MIE}}^{\textup{plug-in}} & =\frac{1}{n}\sum_{i=1}^{n}\frac{\dot{\pi}^{\mathcal{I}_{0}}(Z_{i})}{\sum_{i=1}^{n}\dot{\pi}^{\mathcal{I}_{0}}(Z_{i})/n}\widehat{\textup{MTE}}\big(X_{i},\pi(Z_{i})\big),\label{eq:plug-in}
\end{align}
where $\widehat{\textup{MTE}}\big(x,\pi(z)\big)$ can be evaluated
via Equation \eqref{eq:localIV}, i.e., by taking the partial derivative
of $\mathbb{E}[Y|x,\pi(z)]$ with respect to $\pi(z)$.

In practice, it can be difficult to estimate $\mathbb{E}[Y|x,\pi(z)]$
and its partial derivative nonparametrically, especially when $X$
is high-dimensional. To address this challenge, empirical work using
the MTE often adopts two simplifying assumptions (e.g., \citealt{carneiro2009estimating,Carneiro2011,maestas2013does}).
First, it is typically assumed that $(X,Z)$ is jointly independent
of $(\epsilon,\eta,V)$. This implies that $\textup{MTE}(x,u)$ is
additively separable in $x$ and $u$: 
\begin{align}
\textup{MTE}(x,u) & =\mu_{1}(x)-\mu_{0}(x)+\mathbb{E}[\eta|X=x,U=u]\nonumber \\
 & =\mu_{1}(x)-\mu_{0}(x)+\mathbb{E}[\eta|U=u].\label{eq:mte_additive}
\end{align}
Second, the conditional means of $Y(0)$ and $Y(1)$ given $X$ are
often assumed to be linear in parameters: $\mu_{0}(X)=\beta_{0}^{T}X$
and $\mu_{1}(X)=\beta_{1}^{T}X$. Given the linear specification and
the additive separability of the MTE, $\mathbb{E}[Y|X=x,\pi(Z)=p]$
can be written as
\begin{equation}
\mathbb{E}[Y|X=x,\pi(Z)=p]=\beta_{0}^{T}x+(\beta_{1}-\beta_{0})^{T}xp+\underbrace{\int_{0}^{p}\mathbb{E}[\eta|U=u]du}_{\stackrel{\Delta}{=}K(p)},\label{eq:Yxp2}
\end{equation}
where $K(p)$ is an unknown function of $p$ that can be estimated
either parametrically or nonparametrically.

In the special case where the error terms $(\epsilon,\eta,V)$ are
assumed to be jointly normal with zero means with an unknown covariance
matrix $\Sigma$, the generalized Roy model characterized by Equations
\eqref{eq:Y0}, \eqref{eq:Y1}, \eqref{eq:IA}, and \eqref{eq:A}
is fully parameterized. The unknown parameters $(\beta_{1},\beta_{0},\gamma,\Sigma)$
can then be estimated via maximum likelihood. This model specification
has a long history in econometrics and is commonly referred to as
the ``normal switching regression model'' (\citealt{heckman1978dummy};
see \citealt{Winship1992} for a review). Under the joint normality
assumption, Equation \eqref{eq:mte_additive} reduces to
\begin{equation}
\textup{MTE}(x,u)=(\beta_{1}-\beta_{0})^{T}x+\frac{\sigma_{\eta V}}{\sigma_{V}}\Phi^{-1}(u)\label{eq:mte_normal_est}
\end{equation}
where $\sigma_{\eta V}$ is the covariance between $\eta$ and $V$,
$\sigma_{V}$ is the standard deviation of $V$, and $\Phi^{-1}(\cdot)$
denotes the inverse of the standard normal distribution function.
Substituting the maximum likelihood estimates of $(\beta_{1},\beta_{0},\sigma_{\eta V},\sigma_{V})$
into Equation \eqref{eq:mte_normal_est} yields a parametric estimate
of $\textup{MTE}(x,u)$ for any combination of $x$ and $u$.

The joint normality of error terms is a strong and restrictive assumption.
When the errors $(\epsilon,\eta,V)$ are not normally distributed,
imposing normality can lead to substantial bias in the estimates of
target parameters (\citealt{arabmazar1982investigation}). To address
this problem, \citet{HeckmanUrzuaVytlacil2006a} propose estimating
Equation \eqref{eq:Yxp2} semiparametrically through Robinson's (1988)
partialing-out procedure. In this case, the estimation of $\textup{MTE}(x,u)$
can be summarized in four steps:
\begin{enumerate}
\item \textit{Propensity score estimation}: Estimate the propensity scores
using a standard logit or probit model, and denote them by $\hat{\pi}(Z)$;
\item \textit{Partialing out}: Fit local linear regressions of $Y$, $X$,
and $X\hat{\pi}(Z)$ on $\hat{\pi}(Z)$ and extract the residuals
$e_{Y}$, $e_{X}$, and $e_{X\hat{\pi}(Z)}$;
\item \textit{Parametric component}: Fit a simple linear regression of $e_{Y}$
on $e_{X}$ and $e_{X\hat{\pi}(Z)}$ (without an intercept) to estimate
the parametric component of Equation \eqref{eq:Yxp2}, i.e., $(\beta_{0},\beta_{1}-\beta_{0})$,
and store the residuals of $Y$ from this regression as $e_{Y}^{*}=Y-\hat{\beta}_{0}^{T}X-(\hat{\beta}_{1}-\hat{\beta}_{0})^{T}X\hat{\pi}(Z)$.
\item \textit{Nonparametric component}: Fit a local quadratic regression
(\citealt{fan1996local}) of $e_{Y}^{*}$ on $\hat{\pi}(Z)$ to estimate
$K(p)$ and its derivative $K'(p)$.
\end{enumerate}
A semiparametric estimator of the MTE is then given by
\begin{equation}
\widehat{\textup{MTE}}(x,u)=(\hat{\beta}_{1}-\hat{\beta}_{0})^{T}x+\hat{K}'(u).\label{eq:mte_semi_est}
\end{equation}

Either the parametric or the semiparametric estimator of $\widehat{\textup{MTE}}(x,u)$
can be used to construct a plug-in estimate of the MIE (Equation \ref{eq:plug-in}).
Both approaches, however, rely on correct specification of the outcome
model $\mathbb{E}[Y|X=x,\pi(Z)=p]$. 

\subsubsection{Doubly Robust Estimation}

We now outline a third approach that is more robust to potential misspecification
of the outcome model. Specifically, in settings where $\dot{\pi}^{\mathcal{I}_{0}}(z)$
is a known function of $x$ and $\pi(z)$, we can construct a doubly
robust estimator of the MIE based on the EIF for weighted average
derivatives (\citealt{powell1989semiparametric,newey1993efficiency}).
Define $w(X,\pi(Z))\stackrel{\Delta}{=}\dot{\pi}^{\mathcal{I}_{0}}(Z)/\mathbb{E}[\dot{\pi}^{\mathcal{I}_{0}}(Z)]$,
$m(X,\pi(Z))\stackrel{\Delta}{=}\mathbb{E}[Y|X,\pi(Z)]$, and let
$f(X,\pi(Z))$ denote the density of $(X,\pi(Z))$. The MIE can then
be written as
\begin{equation}
\textup{MIE}=\mathbb{E}[w(X,\pi(Z))\frac{\partial m(X,\pi(Z))}{\partial\pi(Z)}].\label{eq:MIE_functional}
\end{equation}
To facilitate estimation, we follow \citet{newey1993efficiency} by
assuming that $w(X,\pi(Z))f(X,\pi(Z))=0$ at the boundary of the support
of $\pi(Z)$. This assumption is reasonable if the positivity assumption
holds such that no units have a propensity score of zero or one, or
if $w(X,0)=w(X,1)=0$ by construction, such as in the case where $\dot{\pi}^{\mathcal{I}_{0}}(z)=\pi(z)(1-\pi(z))$.
Under this assumption and the assumption that $\pi(Z)$ is known,
the EIF of Equation \eqref{eq:MIE_functional} is (\citealt{newey1993efficiency})
\begin{equation}
\psi(x,\pi(z),y)=w(X,\pi(Z))\frac{\partial m(X,\pi(Z))}{\partial\pi(Z)}+l(X,\pi(Z))[Y-m(X,\pi(Z))]-\textup{MIE},\label{eq:MIE_eif}
\end{equation}
where
\begin{equation}
l(X,\pi(Z))=-\frac{\partial w(X,\pi(Z))}{\partial\pi(Z)}-w(X,\pi(Z))\frac{\partial\textup{log}f(\pi(Z)|X)}{\partial\pi(Z)}.\label{eq:l_function}
\end{equation}
This EIF suggests an estimating equation for the MIE. Suppose we have
fit the following models: $\hat{\pi}(z)$ for $\pi(z)$, $\hat{m}(x,\pi(z))$
for $m(x,\pi(z))$, and $\hat{f}(\pi(z)|x)$ for the conditional density
$f(\pi(z)|x)$. An MIE estimator can then be constructed as
\begin{equation}
\widehat{\textup{MIE}}^{\textup{DR}}=\frac{1}{n}\sum_{i=1}^{n}\big\{ w(X_{i},\hat{\pi}(Z_{i}))\frac{\partial\hat{m}(X_{i},\hat{\pi}(Z_{i}))}{\partial\hat{\pi}(Z_{i})}+\hat{l}(X_{i},\hat{\pi}(Z_{i}))[Y_{i}-m(X_{i},\hat{\pi}(Z_{i}))]\big\},\label{eq:mie_dr}
\end{equation}
where $\hat{l}(X_{i},\hat{\pi}(Z_{i}))$ is obtained by substituting
$\hat{f}(\hat{\pi}(Z_{i})|X_{i})$ into Equation \eqref{eq:l_function}.

Since the EIF (Equation \ref{eq:MIE_eif}) is derived under the assumption
that $\pi(Z)$ is known, whereas in practice it is often estimated,
$\widehat{\textup{MIE}}^{\textup{DR}}$ is not fully efficient. Nonetheless,
it is still doubly robust: if the propensity score model $\pi(z)$
is correctly specified, the estimator remains consistent provided
that if either (a) $m(X_{i},\pi(Z))$ and $\partial m(X,\pi(Z))/\partial\pi(Z)$
are consistently estimated, or (b) $\partial\textup{log}f(\pi(Z)|X)/\partial\pi(Z)$
is consistently estimated (\citealt{chernozhukov2016locally,rothe2019properties}).

In practice, $m(X_{i},\pi(Z))$ and $\partial m(X,\pi(Z))/\partial\pi(Z)$
can be estimated using the parametric or the semiparametric approaches
described above. For the partial derivative $\partial\textup{log}f(\pi(Z)|X)/\partial\pi(Z)$,
one can invoke a location shift model for $\pi(Z)|X$:
\begin{equation}
\pi(Z)=\mathbb{E}[\pi(Z)|X]+\epsilon,\quad\epsilon\ci X.\label{eq:location-shift}
\end{equation}
For this model, we can estimate $\mathbb{E}[\pi(Z)|X]$ using either
a GLM or data-adaptive methods such as Lasso. We can then estimate
$\partial\textup{log}f(\pi(Z)|X)/\partial\pi(Z)$ using a kernel estimator
for the density of $\epsilon$ and its derivative:
\[
\widehat{\frac{\partial\textup{log}f(\pi(Z_{i})|X_{i})}{\partial\pi(Z_{i})}}=\frac{\hat{\phi}'(\hat{\epsilon}_{i})}{\hat{\phi}(\hat{\epsilon}_{i})},
\]
where $\phi(\cdot)$ denotes the density function of $\epsilon$.
If we assume $\epsilon\sim N(0,\sigma_{\epsilon}^{2})$, the above
expression reduces to $-\hat{\epsilon}_{i}/\widehat{\sigma_{\epsilon}^{2}}$.

\subsubsection{Inference}

Because both the plug-in (parametric or semiparametric) and the doubly
robust estimators of the MIE rely on estimated propensity scores $\hat{\pi}(Z_{i})$,
their inference is not straightforward. To account for this additional
layer of uncertainty, it is best to use the nonparametric bootstrap
to estimate their standard errors and confidence intervals.

\subsection{Connections to Existing Work}

The PRTE defined by \citet{PRTE2001} and the MPRTE defined by \citet{Carneiro2010a}
are closely related to the IE and MIE, except that they are conditional
on $X=x$. Analogous to Equations \eqref{eq:IE-MTE} and \eqref{eq:MIE-MTE},
these quantities can be written as weighted means of $\textup{MTE}(x,U)$
and $\textup{MTE}(x,\pi(Z))$, respectively:
\begin{align*}
\textup{PRTE}_{\delta}(x) & =\frac{\mathbb{E}\big[\int_{\pi(Z)}^{\pi^{\mathcal{I}_{\delta}}(Z)}\textup{MTE}\big(X,u\big)du|X=x\big]}{\mathbb{E}\big[\pi^{\mathcal{I}_{\delta}}(Z)-\pi(Z)|X=x\big]},\\
\textup{MPRTE}(x) & =\mathbb{E}\big[\frac{\dot{\pi}^{\mathcal{I}_{0}}(Z)}{\mathbb{E}\big[\dot{\pi}^{\mathcal{I}_{0}}(Z)|X=x\big]}\textup{MTE}\big(X,\pi(Z)\big)|X=x\big].
\end{align*}
To evaluate the overall impact of a marginal policy change, Carneiro
et al. (2010, 2011) propose averaging $\textup{MPRTE}(x)$ over the
marginal distribution of $X$, i.e., $\mathbb{E}[\textup{MPRTE}(X)]$.
As noted in Section \ref{sec:Marginal-Interventional-Effects}, this
estimand can only characterize interventions that vary in intensity
across units different values of $Z\backslash X$ (the excluded instruments)
but not between units with different values of $X$. This is because
the weights linking $\textup{MTE}\big(X,\pi(Z)\big)$ to $\mathbb{E}[\textup{MPRTE}(X)]$
have a mean of one regardless of the covariate values $x$: $\mathbb{E}\big[\dot{\pi}^{\mathcal{I}_{0}}(Z)/\mathbb{E}[\dot{\pi}^{\mathcal{I}_{0}}(Z)|X=x]\big|X=x\big]=1$.
This restriction is unrealistic and undesirable, because real-world
interventions often target individuals and communities defined by
pretreatment covariates such as race, socioeconomic status, or neighborhood
characteristics---variables that serve as covariates rather than
instruments in most empirical applications (see \citealt{zhou2020heterogeneous}
for a more detailed discussion).

By contrast, the MIE is a weighted mean of $\textup{MPRTE}(X)$ (Equation
\ref{eq:MIE-MPRTE}), with the weights proportional to $\dot{\pi}^{\mathcal{I}_{0}}(X)$.
These weights capture the relative intensity of an infinitesimal intervention
across units with covariate values. From this perspective, $\textup{MPRTE}(x)$
can be seen as an intermediate object for constructing the MIE. Nonetheless,
estimation of $\textup{MPRTE}(x)$ is not straightforward, as the
weights linking $\textup{MTE}\big(x,\pi(Z)\big)$ to $\textup{MPRTE}(x)$
involve the conditional density of $\big(\pi(Z),\dot{\pi}^{\mathcal{I}_{0}}(Z)\big)$
given $X=x$. Since $X$ is often high-dimensional, estimation of
these weights is practically challenging and often tackled via ad
hoc methods (e.g., \citealt{Carneiro2011}). By contrast, the plug-in
and doubly robust estimators introduced in Section \ref{subsec:Estimation-IV}
rely directly on Equation \eqref{eq:MIE-MTE}, thereby avoiding the
need to model $f(\pi(Z),\dot{\pi}^{\mathcal{I}_{0}}(Z)|X=x)$.

More recently, \citet{zhou2019marginal,zhou2020heterogeneous} propose
a modified approach to studying marginal policy effects. This approach
starts with a redefinition of the MTE as the expected treatment effect
conditional on the baseline propensity score $\pi(Z)$, rather than
the full vector of pretreatment covariates $X$, together with the
latent resistance to treatment $U$: 
\[
\widetilde{\textup{MTE}}(p,u)\stackrel{\Delta}{=}\mathbb{E}[Y(1)-Y(0)|\pi(Z)=p,U=u].
\]
Here, $\widetilde{\textup{MTE}}(p,u)$ is a bivariate function that
summarizes how treatment effects vary both by observed characteristics
($\pi(Z)$) and by unobserved resistance ($U$). Because a unit is
treated if and only if $\pi(Z)\geq U$, Zhou and Xie define a new
version of the MPRTE:
\[
\widetilde{\textup{MPRTE}}(p)\stackrel{\Delta}{=}\widetilde{\textup{MTE}}(p,p).
\]
$\widetilde{\textup{MPRTE}}(p)$ is a univariate function that reveals
how treatment effects vary by the propensity score $\pi(Z)$ among
those who are exactly indifferent between treatment and non-treatment,
i.e., the units for whom $\pi(Z)=U$. These authors then consider
policy changes of the form $\pi^{\mathcal{I}_{\delta}}(Z)=\pi(Z)+\delta\cdot\lambda(\pi(Z))$,
where $\lambda(\cdot)$ is a known function, and show that the unconditional
MPRTE can be expressed as a weighted average of $\widetilde{\textup{MPRTE}}(p)$:
\begin{align}
\widetilde{\textup{MPRTE}} & =\frac{\mathbb{E}\big[\lambda(\pi(Z))\widetilde{\textup{MPRTE}}(\pi(Z))\big]}{\mathbb{E}\big[\lambda(\pi(Z))\big]}.\label{eq:MPRTEtilde}
\end{align}

Compared with Carneiro et al.'s approach, Zhou and Xie's framework
is \textquotedblleft unconditional\textquotedblright{} in the sense
that it allows the intensity of a policy change to vary with the baseline
propensity score $\pi(Z)$, regardless of whether the variation in
$\pi(Z)$ is driven by the pretreatment covariates $X$ or by the
excluded instruments $(Z\backslash X)$. At the same time, it is more
restrictive in that it only accommodates interventions where the increment
$\pi^{\mathcal{I}_{\delta}}(Z)-\pi(Z)$ is a function of the baseline
propensity score.

In fact, Equation \eqref{eq:MPRTEtilde} follows directly from our
identification result \eqref{eq:MIE-MTE} under the special case $\dot{\pi}^{\mathcal{I}_{0}}(Z)=\lambda(\pi(Z))$:
\begin{align*}
\textup{MIE} & =\mathbb{E}\big[\frac{\dot{\pi}^{\mathcal{I}_{0}}(Z)}{\mathbb{E}\big[\dot{\pi}^{\mathcal{I}_{0}}(Z)\big]}\textup{MTE}\big(X,\pi(Z)\big)\big]\\
 & =\frac{\mathbb{E}[\lambda(\pi(Z))\textup{MTE}\big(X,\pi(Z)\big)]}{\mathbb{E}\big[\lambda(\pi(Z))\big]}\\
 & =\frac{\mathbb{E}\big[\lambda(\pi(Z))\mathbb{E}[\textup{MTE}\big(X,\pi(Z)\big)|\pi(Z)]\big]}{\mathbb{E}\big[\lambda(\pi(Z))\big]}\\
 & =\frac{\mathbb{E}\big[\lambda(\pi(Z))\widetilde{\textup{MPRTE}}(\pi(Z))\big]}{\mathbb{E}\big[\lambda(\pi(Z))\big]}.
\end{align*}
where the last equality uses the fact that $\mathbb{E}[\textup{MTE}\big(X,\pi(Z)\big)|\pi(Z)]=\widetilde{\textup{MTE}}\big(\pi(Z),\pi(Z)\big)$
(see Zhou and Xie 2019). Thus, our identification formula \eqref{eq:MIE-MTE}
can be viewed a generalization of Zhou and Xie's approach, accommodating
a wider class of interventions beyond those restricted to propensity-score--based
increments.

\subsection{Illustration}

We illustrate the MIE estimators described in Section \ref{subsec:Estimation-IV}
by reanalyzing data from \citeauthor{Carneiro2011}'s (2011) study
of the economic returns to college (see also \citealt{zhou2020heterogeneous}).
The sample consists of 1,747 white males at ages 16-22 in 1979, drawn
from the National Longitudinal Survey of Youth (NLSY79). Treatment
($A$) is college attendance, defined by having attained any post-secondary
education by 1991. The treated group consists of 865 individuals,
and the untreated group consists of 882 individuals. The outcome $Y$
is the natural logarithm of hourly wage around 1991, measured as the
average of non-missing hourly wages (deflated to 1983 dollars) reported
between 1989 and 1993. The pretreatment variables ($X$) include linear
and quadratic terms of mother's years of schooling, number of siblings,
the Armed Forces Qualification Test (AFQT) score adjusted by years
of schooling, permanent local log earnings at age 17 (county log earnings
averaged between 1973 and 2000), permanent local unemployment rate
at age 17 (state unemployment rate averaged between 1973 and 2000),
as well as a dummy variable indicating urban residence at age 14 and
cohort dummies. The excluded instruments ($Z\backslash X$) include:
(a) the presence of a four-year college in the county of residence
at age 14; (b) local wage in the county of residence at age 17; (c)
local unemployment rate in the state of residence at age 17; and (d)
average tuition in public four-year colleges in the county of residence
at age 17; and (e) their interactions with mother's years of schooling,
number of siblings, and the adjusted AFQT score. Following Carneiro
et al. (2011), we also include four variables in $X$ but not in $Z$:
years of experience in 1991, years of experience in 1991 squared,
local log earnings in 1991, and local unemployment rate in 1991. Further
details about the dataset and variable construction can be found in
the online appendix of Carneiro et al. (2011).

We consider four stylized interventions akin to those introduced in
Section \ref{subsec:Special-Cases}, except that the baseline and
interventional propensity scores are now the conditional probabilities
of treatment given $Z$ rather than $X$. Thus, the corresponding
local derivatives of the interventional propensity scores are: (a)
$\dot{\pi}^{\mathcal{I}_{0}}(Z)=\mathbb{I}\big(\pi(Z)<1\big)$; (b)
$\dot{\pi}^{\mathcal{I}_{0}}(Z)=\pi(Z)\mathbb{I}\big(\pi(Z)<1\big)$;
(c) $\dot{\pi}^{\mathcal{I}_{0}}(Z)=1-\pi(Z)$; (d) $\dot{\pi}^{\mathcal{I}_{0}}(Z)=\pi(Z)(1-\pi(Z))$.
In particular, intervention (b) is disequalizing in that it nudges
into college more students with relatively high baseline propensity
scores. Conversely, intervention (c) is equalizing, inducing more
``unlikely college-goers'' into higher education.

Similar to our illustration of the MIE under unconfoundedness, we
also conduct two sets of analyses here. First, we assume that positivity
holds, i.e., $\textup{supp}(\pi(X))\subset(0,1)$, so that the indicator
functions in $\dot{\pi}^{\mathcal{I}_{0}}(Z)$ in cases (a) and (b)
can be ignored. Second, we relax positivity to allow some units to
have propensity scores of exactly zero or one. Unlike identification
under unconfoundedness, however, the MIEs for all four stylized interventions
remain identified in this case. This is because the MIE is a weighted
average of $\textup{MTE}\big(X,\pi(Z)\big)$, which, unlike $\textup{CATE}(X)$,
is identified over the full support of $\pi(Z)$, even when it includes
zero or one. In this case, the indicator functions appearing in $\dot{\pi}^{\mathcal{I}_{0}}(Z)$
in cases (a) and (b) cannot be ignored. Since the true propensity
scores are unknown, we cannot know for sure which units violate positivity.
To operationalize this, we set a unit\textquoteright s propensity
score to zero if its estimated score falls below $\min_{i:A_{i}=1}\hat{\pi}(Z_{i})$,
and to one if its estimated propensity score exceeds $\textup{max}_{i:A_{i}=0}\hat{\pi}(Z_{i})$.
The estimated propensity scores $\hat{\pi}(Z_{i})$ are obtained from
a logistic regression of treatment on all covariates and instruments.
In this data, $\min_{i:A_{i}=1}\hat{\pi}(Z_{i})=0.0324$ and $\textup{max}_{i:A_{i}=0}\hat{\pi}(Z_{i})=0.9775$,
leading us to recode 25 untreated units and 42 treated units.

For each of the four interventions, we estimate the MIE using three
methods: a parametric plug-in estimator where the MTE is estimated
using the normal switching regression model (i.e. Equation \ref{eq:mte_normal_est}),
a semiparametric plug-in estimator where the MTE is estimated using
the partialing-out procedure described in Section \ref{subsec:Estimation-IV}
(i.e., Equation \ref{eq:mte_semi_est}), and the doubly robust estimator
\eqref{eq:mie_dr}. For the doubly robust estimator, we assume a location-shift
model for $\pi(Z)|X$ with normal errors, where $\mathbb{E}[\pi(Z)|X]$
is estimated using a generalized linear model with a logistic link.
Standard errors for these estimators are estimated using the nonparametric
bootstrap with 1,000 replications.

\begin{table}
\caption{MIE estimates under four stylized interventions for the NLSY data.\label{tab:Results2}}

\begin{centering}
{\small{}%
\begin{tabular}{l>{\centering}p{1.6cm}>{\centering}p{1.6cm}>{\centering}p{1.6cm}>{\centering}p{1.6cm}>{\centering}p{1.6cm}>{\centering}p{1.6cm}}
\hline 
\multirow{2}{*}{{\small$\pi^{\mathcal{I}_{\delta}}(z)$}} & \multicolumn{2}{c}{parametric plug-in} & \multicolumn{2}{c}{semiparametric plug-in} & \multicolumn{2}{c}{doubly robust}\tabularnewline
\cline{2-7}
 & assuming positivity & relaxing positivity & assuming positivity & relaxing positivity & assuming positivity & relaxing positivity\tabularnewline
\hline 
{\small$\min\{1,\pi(z)+\delta\}$} & 0.259 (0.158) & 0.26 (0.152) & 0.333 (0.173) & 0.328 (0.169) & 0.468 (0.189) & \tabularnewline
{\small$\min\{1,\pi(z)e^{\delta}\}$} & 0.243 (0.206) & 0.245 (0.192) &  0.2 (0.191) & 0.195 (0.183) & 0.251 (0.234) & 0.251 (0.233)\tabularnewline
{\small$1-(1-\pi(z))e^{-\delta}$} & 0.274 (0.128) & 0.274 (0.128) & 0.463 (0.185) & 0.453 (0.184) & 0.681 (0.203) & \tabularnewline
{\small$\frac{e^{\delta}\pi(z)}{1-\pi(z)+e^{\delta}\pi(z)}$} & 0.232 (0.15) & 0.232 (0.15) & 0.302 (0.154) & 0.295 (0.153) & 0.468 (0.191) & 0.464 (0.191)\tabularnewline
\hline 
\end{tabular}}{\small\smallskip{}
}{\small\par}
\par\end{centering}
Note: In the doubly robust estimator, $m(X,\pi(Z))$ and $\partial m(X,\pi(Z))/\partial\pi(Z)$
are estimated using the semiparametric method described in Section
\ref{subsec:Estimation-IV}, and the partial derivative $\partial\textup{log}f(\pi(Z)|X)/\partial\pi(Z)$
is estimated using a location shift model for $\pi(Z)|X$ with normal
errors.
\end{table}

Table \ref{tab:Results2} reports our estimates of the MIE for the
four stylized interventions. The results are highly consistent under
both assumptions---when positivity holds and when it is relaxed.
When positivity is relaxed, however, the doubly robust estimator cannot
be applied to the first and third interventions. This is because the
corresponding weights $w(X,\pi(Z))$ are not zero when $\pi(Z)=0$,
violating the condition that $w(X,\pi(Z))f(X,\pi(Z))=0$ at the boundary
of the support of $\pi(Z)$. 

Across all cases, the estimates point to substantial marginal returns
to college, although the magnitudes differ across the three estimators.
Under the \textquotedblleft neutral\textquotedblright{} intervention,
for example, the semiparametric plug-in estimator yields an MIE of
about 0.33, implying that college attendance raises hourly wages by
approximately 39\% ($e^{0.33}-1=0.39$) for the students induced into
treatment.

The estimated MIE also varies considerably with the form of the policy
change, particularly when using the semiparametric plug-in and doubly
robust methods. Notably, the equalizing intervention ($\pi^{\mathcal{I}_{\delta}}(z)=1-(1-\pi(z))e^{-\delta}$)
produces the highest MIE estimates, suggesting that expanding access
among \textquotedblleft unlikely college-goers\textquotedblright{}
yields especially large wage gains. This finding is consistent with
Zhou and Xie\textquoteright s (2020) analysis of $\widetilde{\textup{MPRTE}}(p)$,
which shows that marginal entrants at the lower end of the baseline
propensity score distribution tend to benefit the most from college
attendance. These results suggest the potential value of targeting
college access initiatives toward disadvantaged groups with low baseline
probabilities of enrollment: Such equalizing interventions not only
expand educational opportunities but may also deliver the greatest
economic returns among those induced into college.

\section{Concluding Remarks}

In this article, we have expounded the concepts of the interventional
effect (IE) and the marginal interventional effect (MIE), defined
as the per capita effect of a treatment intervention on an outcome
of interest and its limit as the intervention size approaches zero.
Compared with conventional causal estimands, the IE and MIE map more
closely onto real-world policy changes, which typically \textquotedblleft nudge\textquotedblright{}
a small segment of the population at or near the margin of participation.
These parameters can be viewed as unconditional counterparts of the
PRTE and MPRTE introduced in the economics literature (Heckman and
Vytlacil 2001, 2005; Carneiro et al. 2010, 2011). Unlike the PRTE
and MPRTE, however, the IE and MIE are defined without reference to
a latent index model and can be identified either under unconfoundedness
or through instrumental variables.

The generality of our framework provides a bridge between the econometrics
literature on policy-relevant treatment effects and the parallel development
of interventional effects in statistics and epidemiology. Under unconfoundedness,
both the IE and MIE can be identified as weighted averages of CATEs;
in particular, the MIE associated with Kennedy\textquoteright s (2019)
incremental propensity score intervention coincides with the ATO parameter,
previously motivated on purely statistical grounds (Li et al. 2018).
Under instrumental-variable assumptions, we showed that the MIE can
be expressed as a weighted average of $\textup{MTE}(X,\pi(Z))$, i.e.,
the treatment effect among individuals with pretreatment covariates
$X$ who are at the margin of treatment ($U=\pi(Z)$). Because $\textup{MTE}\big(x,\pi(z)\big)$
is identified over the entire support of ($X,\pi(Z)$), the MIE remains
identified even when the variation of $\pi(Z)$ given $X$ is limited.
We further discussed several estimation strategies, including parametric
and semiparametric plug-in estimators of the MTE as well as a doubly
robust estimator based on the efficient influence function of weighted
average derivatives.

Our analysis has focused on binary treatments. A growing body of work
has extended interventional approaches to continuous treatments under
unconfoundedness (e.g., \citealt{diaz2012population,diaz2013assessing,haneuse2013estimation,young2014identification,rothenhausler2019incremental,hines2021parameterising}).
For instance, \citet{diaz2012population} consider interventions that
shift the conditional mean of a continuous treatment given pretreatment
covariates, and \citet{rothenhausler2019incremental} define ``incremental
causal effect'' as the limit of one particular type of \citeauthor{diaz2012population}'s
intervention, one in which the shift in treatment is constant across
units. Future research could extend the IE and MIE to this setting,
exploring a richer class of marginal interventions beyond constant
shifts and developing identification and estimation results under
alternative sets of assumptions.

\section*{Acknowledgements}

The authors thank Ang Yu and two reviewers from the Alexander and
Diviya Magaro Peer Pre-Review Program for helpful comments.

\noindent\clearpage\onehalfspacing\bibliographystyle{chicago}
\bibliography{causality_ref}

\end{document}